\documentclass{article}
\usepackage[utf8]{inputenc}
\usepackage[english]{babel}
\usepackage{multicol}
\usepackage{graphicx}  % standard LaTeX graphics tool;
\usepackage{amsmath}   % For getting proper math eqns
\usepackage[compress]{cite}
\usepackage{amssymb}   % Used for getting various symbols eg. gtrsim, lesssim etc.
\usepackage{bm} % For bold math (esp of lower greek letters)
\usepackage{dcolumn}% Align table columns on decimal point
\usepackage{color}
\usepackage{mathrsfs}
\usepackage{amsfonts}
\usepackage{varioref}
\usepackage{textcomp}
\usepackage[parfill]{parskip}
\RequirePackage[colorlinks,citecolor=blue,urlcolor=magenta,linkcolor=blue]{hyperref}
\allowdisplaybreaks
\newcommand{\mypm}{\mathbin{\smash{%
\raisebox{0.35ex}{%
            $\underset{\raisebox{0.5ex}{$\smash -$}}{\smash+}$%
            }%
        }%
    }%
}
\newcommand{\mymp}{\mathbin{\smash{%
\raisebox{0.35ex}{%
            $\underset{\raisebox{0.5ex}{$\smash +$}}{\smash-}$%
            }%
        }%
    }%
}
\def\RN{Reissner-Nordstr\"{o}m}

\def\dS{de Sitter}
\def\SCC{strong cosmic censorship}

\def\CH{Cauchy horizon}
\def\SCCC{strong cosmic censorship conjecture}
\labelformat{section}{Section #1} 
\labelformat{subsection}{Section #1} 
\labelformat{subsubsection}{Section #1}
\labelformat{subsubsubsection}{Section #1}
\labelformat{subfigure}{Fig.~\thefigure#1} 
\labelformat{table}{Table.~#1} 
\labelformat{appendix}{Appendix #1}
\title{On the validity of Strong Cosmic Censorship Conjecture in presence of Dirac Fields}
\author{Mostafizur Rahman\footnote{mostafizur@ctp-jamia.res.in} $^{1}$\\
	{$~^{1}$\small{Center for Theoretical Physics, Jamia Millia Islamia, New Delhi-110025, India}}}
\begin{document}
  
\maketitle
%%%%%%%%%%%%%%%%%%%%%%%%%%%%%%%%%%%%%%%%%%%%%%%%%%%%%%%%%%%%%%%%%%%%%%%%%%%%%%%%%%%%%%%%%%%%%%%%%%%
%%%%%%%%%%%%%%%%%%%%%%%%%%%%%%%%%%%%%%%%%%%%%%%%%%%%%%%%%%%%%%%%%%%%%%%%%%%%%%%%%%%%%%%%%%%%%%%%%%%
%%%%%%%%%%%%%%%%%%%%%%%%%%%%%%%%%%%%%%%%%%%%%%%%%%%%%%%%%%%%%%%%%%%%%%%%%%%%%%%%%%%%%%%%%%%%%%%%%%%

\begin{abstract}
A well posed theory of nature is expected to determine the future of an observer uniquely from a given set of appropriate initial data. In the context of general relativity, this is ensured by Penrose's \SCCC. But in recent years, several examples are found which suggest breakdown of the deterministic nature of the theory in \RN-\dS\ black holes under the influence of different fundamental fields. Nevertheless, the situation has been reassuring for the case of astrophysically meaningful Kerr-\dS\ black hole solutions which seems to respect the conjecture. However, the previous analyses were done considering only the effect of scalar fields. In this paper, we extend the study by considering Dirac fields in Kerr-\dS\ background and show that there exist a parameter space which does not respect the conjecture.
\end{abstract}

\maketitle

\section{Introduction}\label{SCC_Intro}

In recent years, advancement in technology brought a revolution in observational astrophysics that make it possible to test some of most intriguing predictions of general relativity \cite{Abbott:2016blz, Akiyama:2019cqa}. While, due to the uncertainty in the data, the possibility of a modified theory of gravity is not discarded right away, the stature of Einstein's general relativity as the most successful theory of gravity remains unaltered \cite{KONOPLYA2016350, Konoplya:2016hmd,Yunes:2016jcc}. This situation gives confidence to the scientists to propose interesting research work to test the theory in more and more extreme conditions. A possible alternative to this path is to find paradoxes within the theory of general relativity i.e. to check mathematical consistency of the theory. The existence of the Cauchy horizon in Kerr and \RN\ solutions is one of such paradoxes since the theory loses its predictive power beyond that region. However, soon it was realized that the Cauchy horizons are subject to blue shift instability that can turn them into curvature singularities under the influence of even small perturbations \cite{1973IJTP....7..183S, PhysRevD.41.1796, Dafermos:2003wr, Dafermos:2012np}. This phenomena led Penrose to propose the \SCCC\ which can be stated as follows: ``for \textit{a generic initial data}, the maximal Cauchy development (the largest manifold that is uniquely determined by Einstein's field equations from a given set of initial data ) is \textit{ inextendible} as a $C^{0}$ metric'' \cite{Dafermos:2012np,Costa:2017tjc, Costa:2014yha, 0264-9381-16-12A-302,Christodoulou:2008nj}. This conjecture ensures that the observers who dare to cross the Cauchy horizon are torn apart by the infinite tidal forces. An another way to look at the problem is to consider the effect of a linear perturbations on the spacetime metric \cite{PhysRevD.97.104060}. Here, the question of determinism is answered by considering nonlinear effects where the linear perturbations under consideration acts as the source of second order metric perturbations. Adopting the approach mentioned above, a version of \SCCC\ for a Einstein-Maxwell-scalar field system can be stated as follows : ``the maximal Cauchy development of the \textit{stationary, axisymmetric} solutions of a Einstein-Maxwell-scalar field system \textit{can not be extended} across the \CH\ (provided they exist!) with square integrable Christoffel symbols and scalar fields that live on Sobolov space i.e. $\Phi\in H^{1}_{loc}$'' \cite{PhysRevD.97.104060, Dafermos:2018tha, Luk:2015qja,Dafermos:2015bzz}. The same conclusion is obtained for the static and spherically symmetric solutions of a Einstein-Maxwell-Dirac field system with the Dirac field $\Psi$ replacing the scalar field $\Phi$ \cite{Ge:2018vjq, Destounis:2018qnb}. 
\\
Although, any contradiction to this version of \SCCC\ is yet to be found for asymptotically flat black holes \cite{Dafermos:2012np}, the same conclusion may not be drawn in the presence of a positive cosmological constant \cite{Chambers:1997ef}. In this scenario, the effect of blue shift amplification at the \CH\ may be compromised by the exponential decay of the perturbations at event horizon. In fact, recently Cardoso et al. have found a finite parameter space where \SCCC\ gets violated in \RN -\dS\ black holes in presence of a massless, neutral scalar perturbations \cite{Cardoso:2017soq}. Since then, significant amount of works have been done on this topics, further confirming the same for several fundamental fields in a  \RN -\dS\ background \cite{Ge:2018vjq,Destounis:2018qnb,Cardoso:2018nvb, Mo:2018nnu, Rahman:2018oso,PhysRevD.99.064014,Liu:2019lon, Rahman:2020guv}. Interestingly enough, \SCCC\ seems to be always respected in rotating black holes in \dS\ background in the presence of scalar fields \cite{PhysRevD.97.104060, Rahman:2018oso, Rahman:2020guv}. Here, the rotation of the black hole plays a crucial role to uphold the conjecture. So it is legitimate to check whether the conjecture still holds true in presence of other fundamental fields in rotating black hole background in order to better understand the effect of the rotation. In this paper, we study the effect of Dirac fields on \SCC\ in Kerr-\dS\ background. From an astrophysical perspective, this situation is extremely important since Kerr-\dS\ solution describes the most realistic black holes in an expanding universe. 
\\
The paper is organized as follows: In \ref{Weak_solutions}, we constitute weak solutions of Einstein's equations in presence of Dirac field and discuss the effect of these solutions on \SCCC\ which we use to find the the criteria for \SCC\ violation in \ref{sec_Dirac_equation}. In \ref{main_result}, we present the main results of the paper.  Finally, with some relevant discussions in last section, we conclude our paper. Throughout the paper, we use units in which $G = c = \hslash = 1$.
%%%%%%%%%%%%%%%%%%%%%%%%%%%%%%%%%%%%%%%%%%%%%%%%%%%%%%%%%%%%%%%%%%%%%%%%%%%%%%%%%%%%%%%%%%%%%%%%%%%
%%%%%%%%%%%%%%%%%%%%%%%%%%%%%%%%%%%%%%%%%%%%%%%%%%%%%%%%%%%%%%%%%%%%%%%%%%%%%%%%%%%%%%%%%%%%%%%%%%%
%%%%%%%%%%%%%%%%%%%%%%%%%%%%%%%%%%%%%%%%%%%%%%%%%%%%%%%%%%%%%%%%%%%%%%%%%%%%%%%%%%%%%%%%%%%%%%%%%%%
\section{Weak solution of Einstein Equation in presence of massless Dirac fields}\label{Weak_solutions}

The fate of \SCCC\ relies on the possibility of finding a solution of Einstein equation at the \CH. Even if the metric is not differentiable (but continuous!) at the \CH, one can still make sense of Einstein equation there by constituting a \textit{weak}\ solution of the equation \cite{PhysRevD.97.104060}. This can be understood by considering the effect of linear perturbation on the spacetime. Consider a massless Dirac field which satisfies the equation $\widehat{\mathcal{D}}\Psi=0$, triggers a perturbation in the spacetime. Here $\widehat{\mathcal{D}}$ is the Dirac operator that acts on the spinor $\Psi$. Let the Dirac field act as a first order perturbation which induces a second order perturbation of the metric, denoted by $h_{\mu\nu}^{(2)}$ which satisfies the following equation
%%%%%%%%%%%%%%%%%%%%%%%%%%%%%%%%%%%%%%%%%%%%%%%%%%%%
\begin{equation}\label{weak_Pertubation}
\widehat{\mathcal{O}}h_{\mu\nu}^{(2)}=8\pi~T_{\mu\nu}^{\Psi}
\end{equation}
%%%%%%%%%%%%%%%%%%%%%%%%%%%%%%%%%%%%%%%%%%%%%%%%%%%%
where, $\widehat{\mathcal{O}}$ is a second order differential operator  and $T_{\mu\nu}^{\Psi}$ is the stress-energy tensor for the Dirac field which can be expressed in the following form
%%%%%%%%%%%%%%%%%%%%%%%%%%%%%%%%%%%%%%%%%%%%%%%%%%%%
\begin{equation}\label{EM_tensor}
\begin{aligned}
T_{\mu\nu}^{\Psi}&=~\frac{i}{2}~\left[\overline{\Psi}\gamma_{(\mu}\nabla_{\nu)}\Psi-\nabla_{(\mu}\overline{\Psi}\gamma_{\nu)}\Psi\right]\\\nonumber&-~\frac{i}{2}~g_{\mu\nu}\left[\overline{\Psi}\gamma^{\lambda}\nabla_{\lambda}\Psi-\nabla_{\lambda}\overline{\Psi}\gamma^{\lambda}\Psi\right].
\end{aligned}
\end{equation}
%%%%%%%%%%%%%%%%%%%%%%%%%%%%%%%%%%%%%%%%%%%%%%%%%%%% 
Even when $h_{\mu\nu}^{(2)}$ is not differentiable at Cauchy horizon, we can have a solution of Einstein equation by multiplying Eq.~(\ref{weak_Pertubation}) with a smooth, symmetric tensor $K^{\mu\nu}$. By performing integration by parts, we obtain the following equation
%%%%%%%%%%%%%%%%%%%%%%%%%%%%%%%%%%%%%%%%%%%%%%%%%%%%
\begin{equation}\label{int_weak_sol}
\int_{\mathcal{M}}~d^{4}x~\sqrt{-g}\left(h_{\mu\nu}^{(2)}~\mathcal{L}^{\dagger}~K^{\mu\nu}\right)=\int_{\mathcal{M}}~d^{4}x~\sqrt{-g}\left(K^{\mu\nu}~T_{\mu\nu}^{\Psi}\right)
\end{equation}
%%%%%%%%%%%%%%%%%%%%%%%%%%%%%%%%%%%%%%%%%%%%%%%%%%%%
where, $\mathcal{M}$ is the \textit{maximal Cauchy development} i.e. the region of spacetime uniquely determined by a set of generic initial data \cite{Dafermos:2018tha, Costa:2017tjc} and  $\mathcal{L}^{\dagger}$ is the adjoint of the operator $\mathcal{L}$. If this equation is satisfied for any smooth, symmetric function $K^{\mu\nu}$, there exists a  \textit{weak} solution of Einstein equation provided that both sides of the equation remains finite. The requirement is fulfilled  when $\Psi$ belongs to $H^{1}_{loc}$. When such solutions exists at the \CH, we can extend the metric across it which leads to the breakdown of \SCCC.

%%%%%%%%%%%%%%%%%%%%%%%%%%%%%%%%%%%%%%%%%%%%%%%%%%%%%%%%%%%%%%%%%%%%%%%%%%%%%%%%%%%%%%%%%%%%%%%%%%
%%%%%%%%%%%%%%%%%%%%%%%%%%%%%%%%%%%%%%%%%%%%%%%%%%%%%%%%%%%%%%%%%%%%%%%%%%%%%%%%%%%%%%%%%%%%%%%%%%
\section{Dirac equation in Kerr-de Sitter spacetime and the criteria of violation of Strong Cosmic Censorship Conjecture}\label{sec_Dirac_equation}

In the previous section, we have found that the presence of Dirac fields in Kerr-\dS\ spacetimes can lead to the violation of \SCC, if the spinor field $\Psi$ belongs to $H^{1}_{loc}$ at \CH. In this section, we closely inspect this condition. We solve the Dirac equation near the \CH\ of Kerr-\dS\ black holes and then rewrite the condition in terms of black hole parameters. We start with  a Kerr-\dS\ black hole spacetime in Boyer-Lindquist coordinate $(t,r,\theta,\phi)$ whose line element can be expressed as follows \cite{Suzuki:1998vy}
%%%%%%%%%%%%%%%%%%%%%%%%%%%%%%%%%%%%%%%%%%%%%%%%%%%%
\begin{eqnarray}\label{KDS_metric}
ds^{2}=-\frac{\Delta_{r}}{(1+\alpha)^{2}\Sigma}\left[dt-a \sin^{2}\theta~d\phi\right]^{2}+\Sigma\left[\frac{dr^{2}}{\Delta_{r}}+\frac{d\theta^{2}}{\Delta_{\theta}}\right]\nonumber\\
+\frac{\Delta_{\theta}\sin^{2}\theta}{(1+\alpha)^{2}\Sigma}\left[a dt-(r^{2}+a^{2})~d\phi\right]^{2}\nonumber\\
\end{eqnarray}
%%%%%%%%%%%%%%%%%%%%%%%%%%%%%%%%%%%%%%%%%%%%%%%%%%%%
where,
%%%%%%%%%%%%%%%%%%%%%%%%%%%%%%%%%%%%%%%%%%%%%%%%%%%%
\begin{eqnarray}
& {\Delta_{r}}(r)=(r^{2}+a^{2})(1-\frac{\Lambda r^{2}}{3})-2M r\,,\qquad 
{\Delta_{\theta}}(\theta)=1+\frac{\Lambda a^{2}}{3}\cos^{2}\theta\,, \nonumber \\
& {\alpha}=\frac{\Lambda a^{2}}{3}
\,,\qquad 
{\varrho}=r+i a\cos\theta\ \nonumber 
\,,\qquad 
{\Sigma}=\varrho\varrho^{*}\ \nonumber 
\end{eqnarray}
%%%%%%%%%%%%%%%%%%%%%%%%%%%%%%%%%%%%%%%%%%%%%%%
Here, $M$ is the mass of the black hole, $a$ is the black hole rotation parameter and $\Lambda>0$ is the cosmological constant. Throughout the paper, superscript `$*$' denotes the complex conjugate of a quantity. Since we want to check the validity of strong cosmic censorship conjecture in presence of positive cosmological constant, we choose the values of the black hole parameters $M$, $a$ and $\Lambda$ in such a way that the spacetime possesses three distinct horizons.  The position of Cauchy, event and cosmological horizon which we denote by $r_{-}$, $r_{+}$ and $r_{c}$ respectively, can be found by solving the equation $\Delta_{r}(r)=0$. The properties of Dirac fields in this spacetime can be best understood in the framework of Newman-Penrose formalism \cite{Chandrasekhar:1985kt, Frolov:1998wf}. Here, we choose our null tetrad  to be \cite{Suzuki:1998vy,PhysRevD.28.1291,Chang:2005ki}
%%%%%%%%%%%%%%%%%%%%%%%%%%%%%%%%%%%%%%%%%%%%%%%%%%%%
\begin{equation}\label{tetrad_t}
\begin{aligned}
l^{\mu}&=[\dfrac{(1+\alpha)(r^2+a^2)}{\Delta_{r}},1,0,\frac{(1+\alpha)a}{\Delta_{r}}]\\
%%%%%%%%
n^{\mu}&=\frac{1}{2\Sigma}[(1+\alpha)(r^2+a^2),-\Delta_{r},0,(1+\alpha)a]\\
%%%%%%%
m^{\mu}&=\frac{1}{\sqrt{2\Delta_{\theta}}\varrho}[i a (1+\alpha)\sin\theta,0,\Delta_{\theta},\frac{i(1+\alpha)}{\sin\theta}]\\
%%%%%%%
\bar{m}^{\mu}&=m^{*\mu}
\end{aligned}
\end{equation}
%%%%%%%%%%%%%%%%%%%%%%%%%%%%%%%%%%%%%%%%%%%%%%% 
in the $(t,r,\theta,\phi)$ coordinate. We can easily verify that the only non-vanishing inner product combination is given by the normalization condition, $\bold{l}\cdot\bold{n}=-1$ and $\bold{m}\cdot\bold{\bar{m}}=1$. The advantage of this choice is that the tetrad vectors are regular across the Cauchy horizon. To see that we write the tetrad in outgoing Eddington-Finkelstein coordinate $(u,r,\theta,\varphi)$ by using the following transformation \cite{Zannias:2017ign}
%%%%%%%%%%%%%%%%%%%%%%%%%%%%%%%%%%%%%%%%%%%%%%%%%%%%%%%%%%%%%%%%%%%%%%%%%%%%%%%%%%%%%%%%%%%%%%%%%%
%%%%%%%%%%%%%%%%%%%%%%%%%%%%%%%%%%%%%%%%%%%%%%%%%%%%%%%%%%%%%%%%%%%%%%%%%%%%%%%%%%%%%%%%%%%%%%%%%%
\begin{eqnarray}
& &{du}=dt-\frac{(1+\alpha)(r^{2}+a^{2})}{\Delta_{r}}dr\,,\qquad 
{d\varphi}=d\phi-\frac{(1+\alpha)a}{\Delta_{r}}dr \nonumber 
\end{eqnarray} 
%%%%%%%%%%%%%%%%%%%%%%%%%%%%%%%%%%%%%%%%%%%%%%%%%%%%%%%%%%%%%%%%%%%%%%%%%%%%%%%%%%%%%%%%%%%%%%%%%%
Under this transformation the tetrad vectors take the following form
%%%%%%%%%%%%%%%%%%%%%%%%%%%%%%%%%%%%%%%%%%%%%%%%%%%%%%%%%%%%%%%%%%%%%%%%%%%%%%%%%%%%%%%%%%%%%%%%%%
\begin{equation}\label{tetrad_u}
\begin{aligned}
l^{\mu}&=[0,1,0,0]\\
%%%%%%%%
n^{\mu}&=\frac{1}{\Sigma}[(1+\alpha)(r^2+a^2),\frac{-\Delta_{r}}{2},0,(1+\alpha)a]\\
%%%%%%%
m^{\mu}&=\frac{1}{\sqrt{2\Delta_{\theta}}\varrho}[i a (1+\alpha)\sin\theta,0,\Delta_{\theta},\frac{i(1+\alpha)}{\sin\theta}]\\
%%%%%%%
\bar{m}^{\mu}&=m^{*\mu}.\\
%%%%%%%
\end{aligned}
\end{equation}
%%%%%%%%%%%%%%%%%%%%%%%%%%%%%%%%%%%%%%%%%%%%%%%%%%%%%%%%%%%%%%%%%%%%%%%%%%%%%%%%%%%%%%%%%%%%%%%%%%
Note that, the tetrad vectors are regular at the Cauchy horizon. \\
%%%%%%%%%%%%%%%%%%%%%%%%%%%%%%%%%%%%%%%%%%%%%%%%%%%%%%%%%%%%%%%%%%%%%%%%%%%%%%%%%%%%%%%%%%%%%%%%%%
%%%%%%%%%%%%%%%%%%%%%%%%%%%%%%%%%%%%%%%%%%%%%%%%%%%%%%%%%%%%%%%%%%%%%%%%%%%%%%%%%%%%%%%%%%%%%%%%%%
%%%%%%%%%%%%%%%%%%%%%%%%%%%%%%%%%%%%%%%%%%%%%%%%%%%%%%%%%%%%%%%%%%%%%%%%%%%%%%%%%%%%%%%%%%%%%%%%%%
In Newman-Penrose formalism, the equation for a massless Dirac field $\Psi$ can be written as four coupled differential equations as follows \cite{Chandrasekhar:1985kt}
%%%%%%%%%%%%%%%%%%%%%%%%%%%%%%%%%%%%%%%%%%%%%%%%%%%%%%%%%%%%%%%%%%%%%%%%%%%%%%%%%%%%%%%%%%%%%%%%%%
\begin{equation}\label{dirac}
\begin{aligned}
(D+\epsilon-\rho)F_{1}+(\bar{\delta}+\pi-\alpha)F_{2}&=0\\
%%%%%%%%
(\Delta+\mu-\gamma)F_{2}+(\delta+\beta-\tau)F_{1}&=0\\
%%%%%%%
(D+\epsilon^{*}-\rho^{*})G_{2}-(\delta+\pi^{*}-\alpha^{*})G_{1}&=0\\
%%%%%%%
(\Delta+\mu^{*}-\gamma^{*})G_{1}-(\bar{\delta}+\beta^{*}-\tau^{*})G_{2}&=0.\\
%%%%%%%
\end{aligned}
\end{equation}
%%%%%%%%%%%%%%%%%%%%%%%%%%%%%%%%%%%%%%%%%%%%%%%%%%%%%%%%%%%%%%%%%%%%%%%%%%%%%%%%%%%%%%%%%%%%%%%%%%
where, $D=l\cdot\nabla$, $\Delta=n\cdot\nabla$, $\delta=m\cdot\nabla$, $\bar\delta=\bar m\cdot\nabla$ are the directional covariant derivative along the tetrad vectors and $\alpha$, $\beta$, $\gamma$, $\epsilon$, $\pi$, $\rho$ and $\tau$ are the spin coefficients (for details see \cite{Chandrasekhar:1985kt, Frolov:1998wf}). Here, $F_{1}$, $F_{2}$, $G_1$, $G_2$ denote the spinor components such that $\Psi=(F_1,F_2,-G_2,G_1)^{T}$. For our choice of tetrad (given by Eq.~(\ref{tetrad_t})), the non-vanishing spin coefficients are given by \cite{PhysRevD.28.1291,Chang:2005ki}
%%%%%%%%%%%%%%%%%%%%%%%%%%%%%%%%%%%%%%%%%%%%%%%%%%%%%%%%%%%%%%%%%%%%%%%%%%%%%%%%%%%%%%%%%%%%%%%%%%
\begin{eqnarray}\label{spin_coef}
{\rho}&=&-\frac{1}{\varrho^{*}}
\,,\qquad 
{\tau}=-\frac{i~a~\sqrt{\widetilde{\Delta_{\theta}}}}{\sqrt{2}\varrho^{2}}
\,,\qquad 
{\pi}=\frac{i~a~\sqrt{\widetilde{\Delta_{\theta}}}}{\sqrt{2}(\varrho^{*})^{2}},\nonumber\\
%%%%%%%%%%%%%%%%%%%%%%
{\mu}&=&-\frac{\Delta_{r}}{2\Sigma\varrho^{*}}
\,,\qquad 
{\gamma}=\mu+\frac{1}{4\Sigma}\frac{d\Delta_{r}}{dr},\nonumber\\
%%%%%%%%%%%%%%%%%%%%%%%%%% 
{\beta}&=&\frac{1}{2\sqrt{2}\rho\sin\theta}\frac{d\sqrt{\widetilde{\Delta}_{\theta}}}{d\theta}
\,,\qquad 
{\alpha}=\pi-\beta^{*}.\nonumber
\end{eqnarray} 
%%%%%%%%%%%%%%%%%%%%%%%%%%%%%%%%%%%%%%%%%%%%%%%%%%%%%%%%%%%%%%%%%%%%%%%%%%%%%%%%%%%%%%%%%%%%%%%%%%
where, $\widetilde{\Delta_{\theta}}=\Delta_{\theta}\sin^{2}\theta$. Due to presence of timelike and angular Killing vectors, the Dirac field can be decomposed as $\Psi(t,r,\theta,\phi)=e^{-i\omega t}e^{i m \phi}(f_{1}, f_{2},-g_{2},g_{1})^{T}$, where $f_{1}$, $f_{2}$, $g_1$, $g_2$ are functions of $r$ and $\theta$ only. Moreover, if   we take the following transformation \cite{Chandrasekhar:1985kt,Chang:2005ki}
%%%%%%%%%%%%%%%%%%%%%%%%%%%%%%%%%%%%%%%%%%%%%%%%%%%%%%%%%%%%%%%%%%%%%%%%%%%%%%%%%%%%%%%%%%%%%%%%%%
\begin{eqnarray}
{f_{1}(r,\theta)}=\frac{R_{-}(r)~S_{-}(\theta)}{\varrho^{*}}\,,\qquad 
{f_{2}(r,\theta)}=R_{+}(r)~S_{+}(\theta)\nonumber \\
%%%%%%%%%%%%%%%%%%%%%%%%%
{g_{1}(r,\theta)}=R_{+}(r)~S_{-}(\theta)\,,\qquad 
{g_{2}(r,\theta)}=\frac{R_{-}(r)~S_{+}(\theta)}{\varrho}\nonumber 
\end{eqnarray} 
%%%%%%%%%%%%%%%%%%%%%%%%%%%%%%%%%%%%%%%%%%%%%%%%%%%%%%%%%%%%%%%%%%%%%%%%%%%%%%%%%%%%%%%%%%%%%%%%%%
Eq.~(\ref{dirac}) can be decomposed into radial and angular parts which can be written as follows \cite{PhysRevD.28.1291,Chang:2005ki}
%%%%%%%%%%%%%%%%%%%%%%%%%%%%%%%%%%%%%%%%%%%%%%%%%%%%
%%%%%%%%%%%%%%%%%%%%%%%%%%%%%%%%%%%%%%%%%%%%%%%%%%%%%%%%%%%%%%%%%%%%%%%%%%%%%%%%%%%%%%%%%%%%%%%%%%
\begin{eqnarray}\label{Rad_eqn}
& {\mathcal{D}_{-}R_{-}}=\lambda \sqrt{\Delta_{r}}R_{+}\,,\qquad 
{\mathcal{D}_{+}\sqrt{\Delta_{r}}R_{+}}=\lambda R_{-} 
\end{eqnarray} 
%%%%%%%%%%%%%%%%%%%%%%%%%%%%%%%%%%%%%%%%%%%%%%%%%%%%%%%%%%%%%%%%%%%%%%%%%%%%%%%%%%%%%%%%%%%%%%%%%%
\begin{eqnarray}\label{Ang_eqn}
& {\sqrt{\Delta_{\theta}}\mathcal{L}_{-}S_{-}}=\lambda S_{+}\,,\qquad 
{\sqrt{\Delta_{\theta}}\mathcal{L}_{+}S_{+}}=-\lambda S_{+} 
\end{eqnarray} 
%%%%%%%%%%%%%%%%%%%%%%%%%%%%%%%%%%%%%%%%%%%%%%%%%%%%%%%%%%%%%%%%%%%%%%%%%%%%%%%%%%%%%%%%%%%%%%%%%%
where,
\begin{equation}
\begin{aligned}
\mathcal{D}_{\mypm}&=\left(\sqrt{\Delta_{r}}\partial_{r}\mymp i\frac{V(r)}{\sqrt{\Delta_{r}}}\right)\\
%%%%%%%%%%%%%%%%%%%%%%%%%%
\mathcal{L}_{\mypm}&=\partial_{\theta}\mymp\frac{1+\alpha}{\Delta_{\theta}}H(\theta)+\frac{1}{2\sqrt{\widetilde{\Delta_{\theta}}}}\dfrac{d\sqrt{\widetilde{\Delta_{\theta}}}}{d\theta}\nonumber\\
%%%%%%%%%%%%%%%%%%%%%%%%%%
\end{aligned}
\end{equation}
%%%%%%%%%%%%%%%%%%%%%%%%%%%%%%%%%%%%%%%%%%%%%%%%%%%%
Here, $\lambda$ is a constant of separation, $V(r)=(1+\alpha)[\omega(r^{2}+a^{2})-a m]$ and $H(\theta)=(a \omega \sin\theta-m\csc\theta)$. 
\\
Since we are interested in the behavior of the Dirac field near the Cauchy horizon, it is convenient to adopt the  outgoing Eddington-Finkelstein coordinate $(u,r,\theta,\varphi)$ \cite{Zannias:2017ign}.  Moreover, if we set $\widetilde{R}_{+}(r)=(\Delta_{r})^{\frac{1}{2}}R_{+}(r)$ and  $\widetilde{R}_{-}(r)=R_{-}(r)$, Eq.~(\ref{Rad_eqn}) can be written in a more symmetric form $\mathcal{D}_{\mypm}\widetilde{R}_{\mypm}=\lambda \widetilde{R}_{\mymp}$. Near the Cauchy horizon, the radial equations can be written as follows
%%%%%%%%%%%%%%%%%%%%%%%%%%%%%%%%%%%%%%%%%%%%%%%%%%%%
\begin{equation}
\dfrac{d^{2}\widehat{R}_{\mypm}}{dr_{*}^{2}}+2 i(\omega-m\Omega_{-})\dfrac{d\widehat{R}_{\mypm}}{dr_{*}}=0
\end{equation}
%%%%%%%%%%%%%%%%%%%%%%%%%%%%%%%%%%%%%%%%%%%%%%%%%%%%
where, $\widehat{R}_{\mypm}=\widetilde{R}_{\mypm}\exp[-i(\omega r_{*}-m r_{\phi})]$ and  $\Omega_{-}$ corresponds to the angular velocity of the black hole at the Cauchy horizon. Here, $dr_{*}=(1+\alpha)(r^{2}+a^2)dr/\Delta_{r}$ and $dr_{\phi}=(1+\alpha)a~dr/\Delta_{r}$. The above equations have two independent solutions such that the spinor field can be written as follows
%%%%%%%%%%%%%%%%%%%%%%%%%%%%%%%%%%%%%%%%%%%%%%%%%%%%
\begin{equation}
\begin{aligned}
\widehat{\Psi}^{(1)}(u,r,\theta,\varphi)&=e^{-i(\omega u- m\varphi)}~\psi(r,\theta)\\
%%%%%%%%
\widehat{\Psi}^{(2)}(u,r,\theta,\varphi)&=e^{-i(\omega u- m\varphi)}~\psi(r,\theta)~(r-r_{-})^{p}
\end{aligned}
\end{equation}
%%%%%%%%%%%%%%%%%%%%%%%%%%%%%%%%%%%%%%%%%%%%%%%%%%%%
where, $\psi=(\frac{\mathcal{R}_{-}~S_{-}}{\varrho^{*}},\mathcal{R}_{+}~S_{+},-\frac{\mathcal{R}_{-}~S_{+}}{\varrho},\mathcal{R}_{+}~S_{-})^{T}$. Here, $\mathcal{R}_{\mypm}(r)$ represent some smooth functions of $r$ which are non-vanishing at the Cauchy horizon and $p=i(\omega-m \Omega_{-})/\kappa_{-}$ where $\kappa_{-}$ corresponds to the surface gravity  of the black hole at the Cauchy horizon. Given the solution of Dirac equation near the Cauchy horizon, we need to check whether the Dirac field belongs to $H^{1}_{loc}$ or not in order to investigate the possibility of having weak solutions of Einstein equations. In other words, we have to check whether $\partial_{\mu} \Psi$ is locally square integrable which boils down to investigate the finiteness of the integral of quantity $\sim (r-r_{-})^{2(p-1)}$. Thus, the condition for $\Psi$ to remain in $H^{1}_{loc}$ then reduces to the following inequality 
%%%%%%%%%%%%%%%%%%%%%%%%%%%%%%%%%%%%%%%%%%%%%%%%%%%%
\begin{center}
	\begin{equation}\label{SCC_violation}
	\beta\equiv-\frac{\operatorname{\mathbb{I}m}(\omega)}{\kappa_{-}}>\frac{1}{2}
	\end{equation}
\end{center}
%%%%%%%%%%%%%%%%%%%%%%%%%%%%%%%%%%%%%%%%%%%%%%%%%%%%
The existence of weak solutions of Einstein's equation at \CH\ is guaranteed by this condition which leads  to  a possible  violation  of  strong  cosmic  censorship  conjecture. In our study, we need to focus on the dominant mode contributions only which are the least damped modes of the quasinormal spectrum.
%%%%%%%%%%%%%%%%%%%%%%%%%
%%%%%%%%%%%%%%%%%%%%%%%%%%%%%%%%%%%%%%%%%%%%%%%%%%%%%%%%%%%%%%%%%%%%%%%%%%%%%%%%%%%%%%%%%%%%%%%%%%%
%%%%%%%%%%%%%%%%%%%%%%%%%%%%%%%%%%%%%%%%%%%%%%%
\begin{figure*}
	%%%%%%%%%%%%%%%%%%%%%%%%
	\minipage{0.33\textwidth}
	\includegraphics[width=\linewidth]{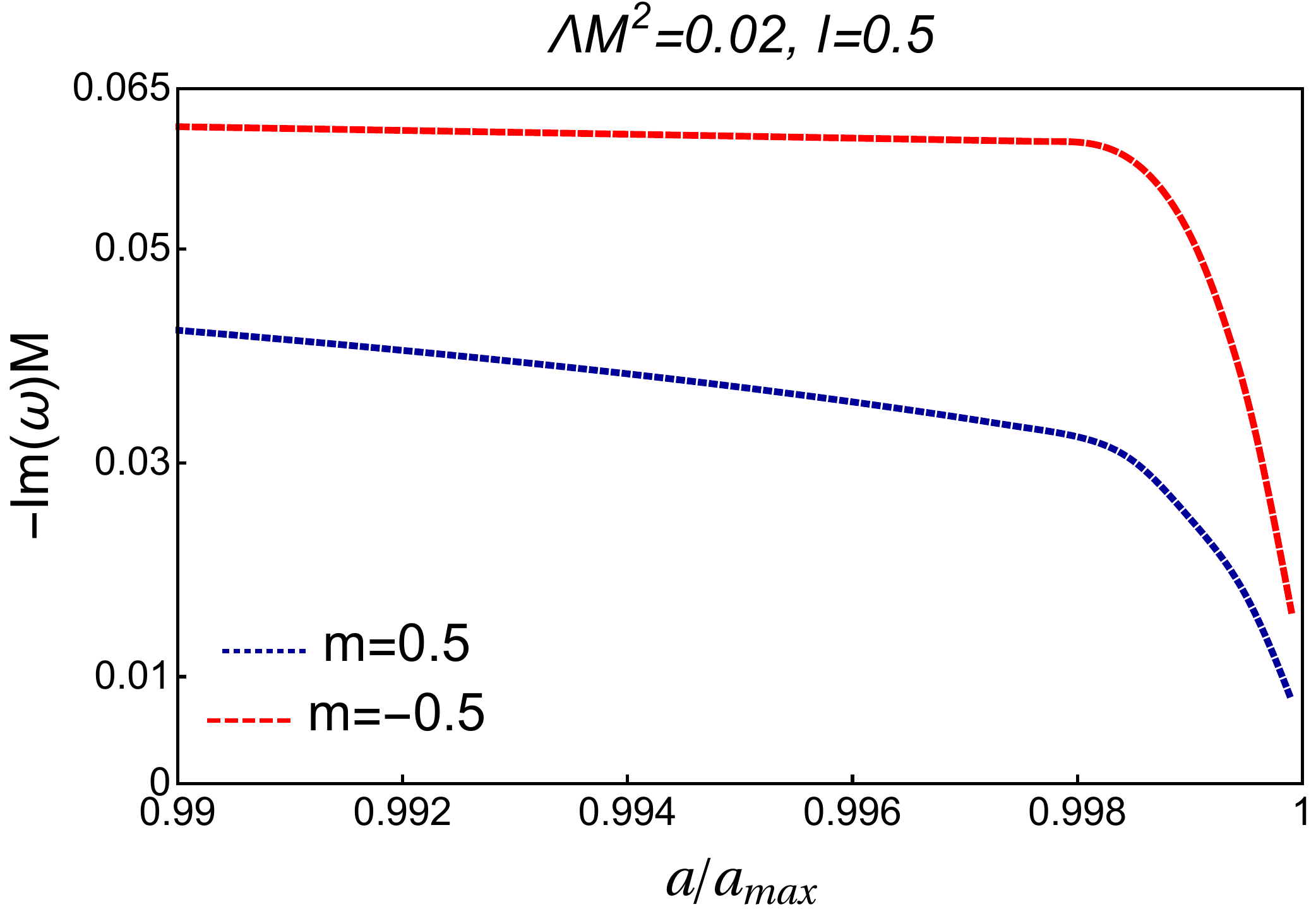}
	\endminipage\hfill
	%%%%%%%%%%%%%%%%%%%%%%%%
	\minipage{0.33\textwidth}
	\includegraphics[width=\linewidth]{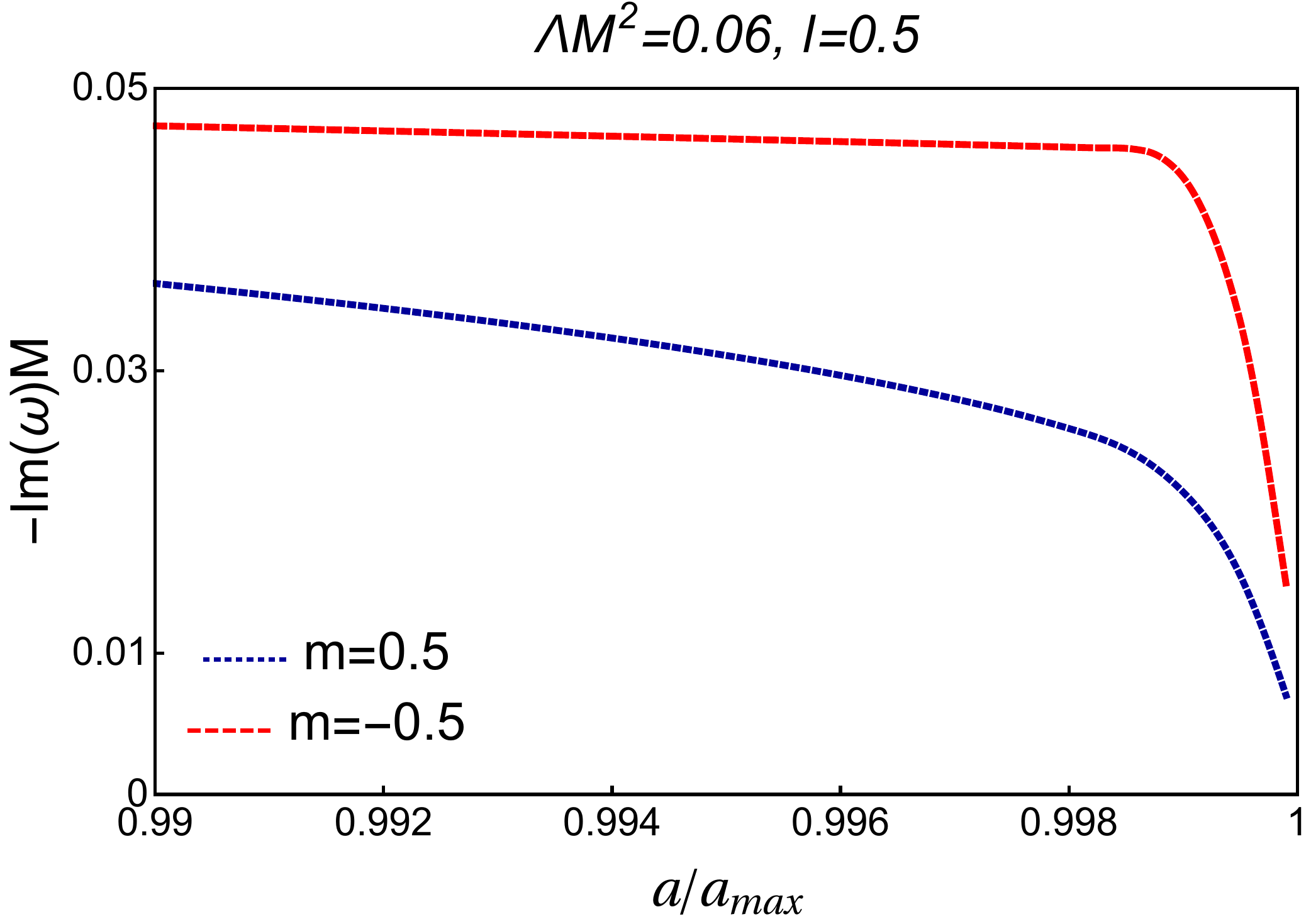}
	\endminipage\hfill
	%%%%%%%%%%%%%%%%%%%%%%%%
	\minipage{0.33\textwidth}%
	\includegraphics[width=\linewidth]{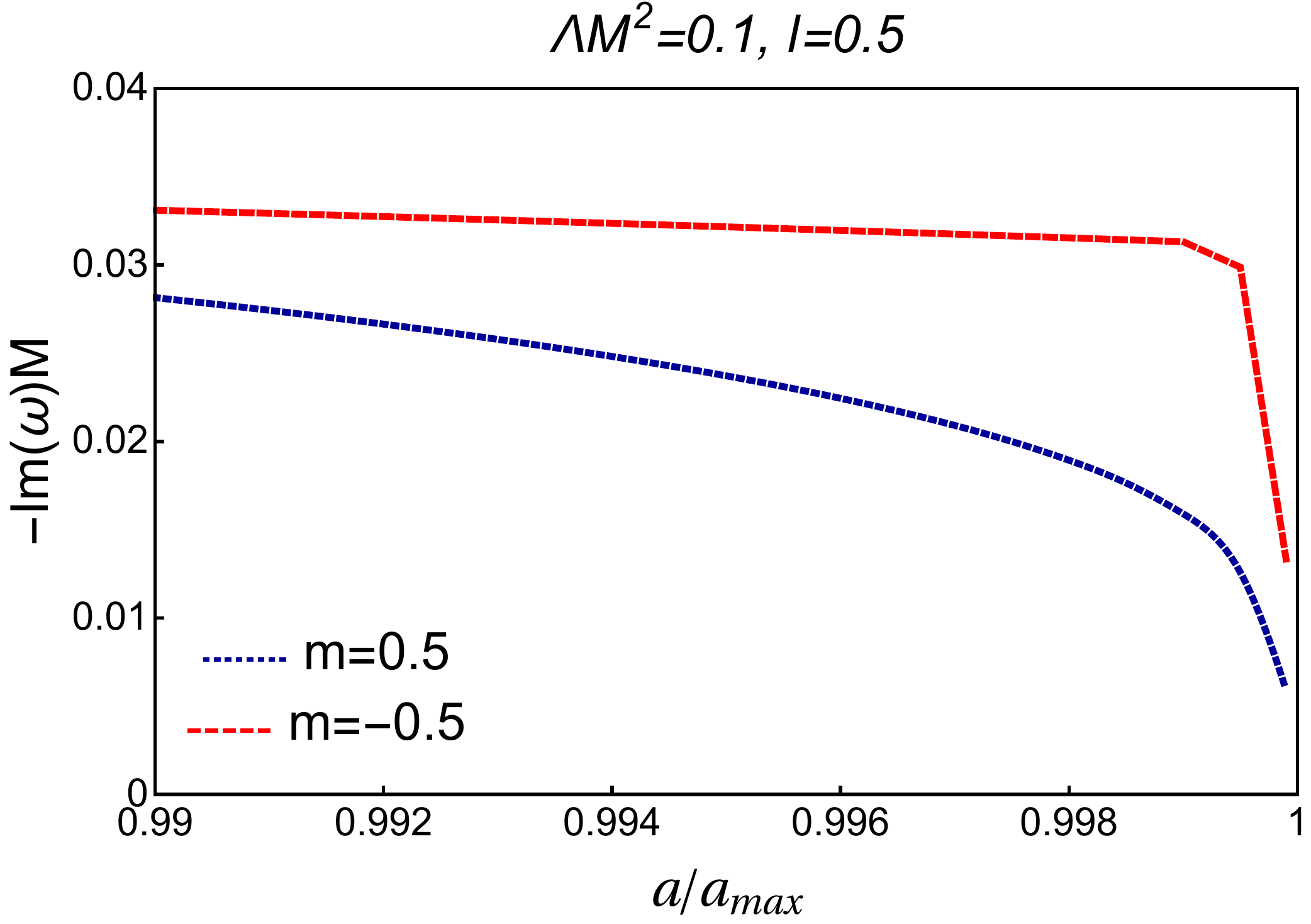}
	\endminipage\hfill
	%%%%%%%%%%%%%%%%%%%%%%%%%
	\minipage{0.33\textwidth}
	\includegraphics[width=\linewidth]{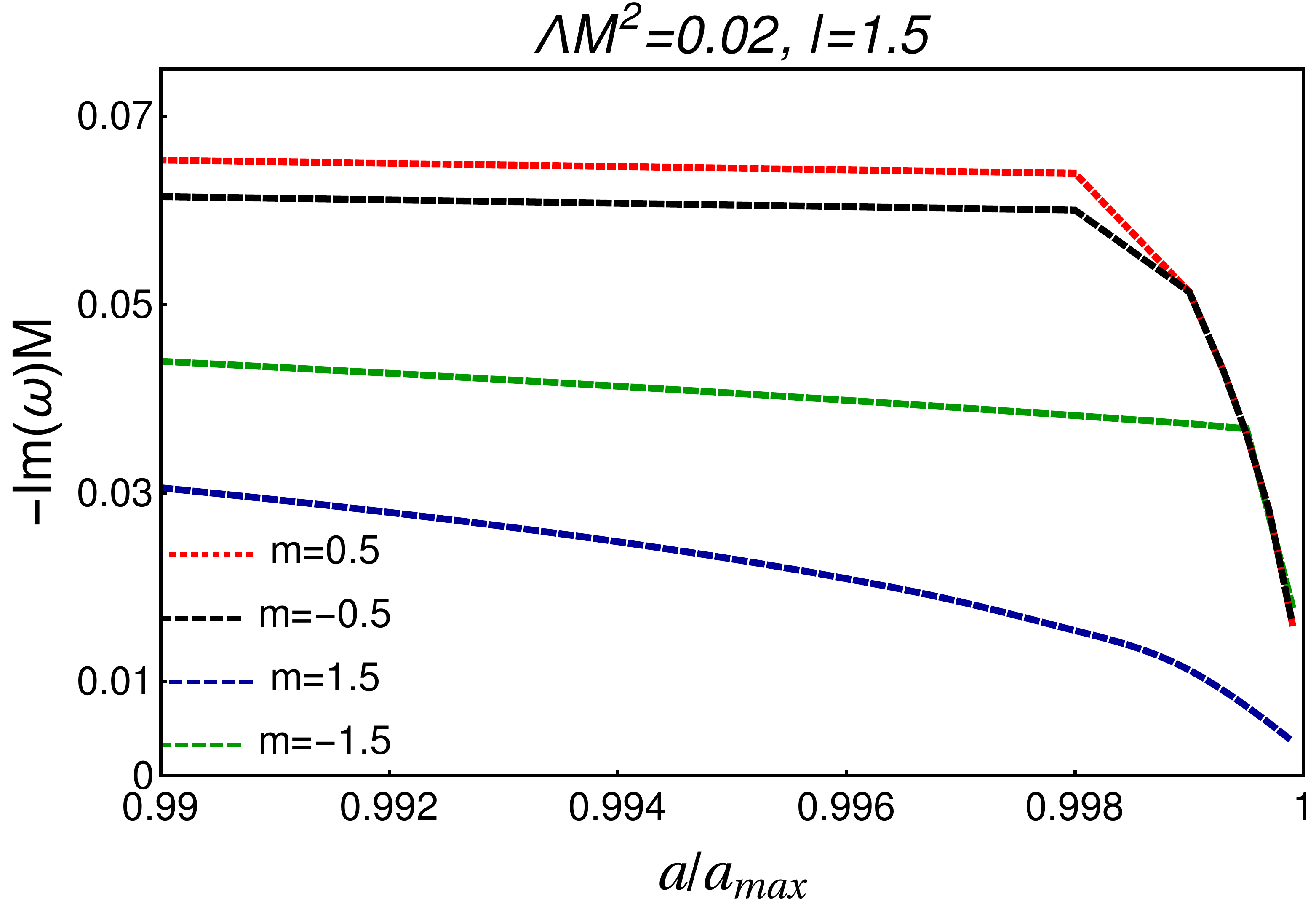}
	\endminipage\hfill
	%%%%%%%%%%%%%%%%%%%%%%%%%
	\minipage{0.33\textwidth}
	\includegraphics[width=\linewidth]{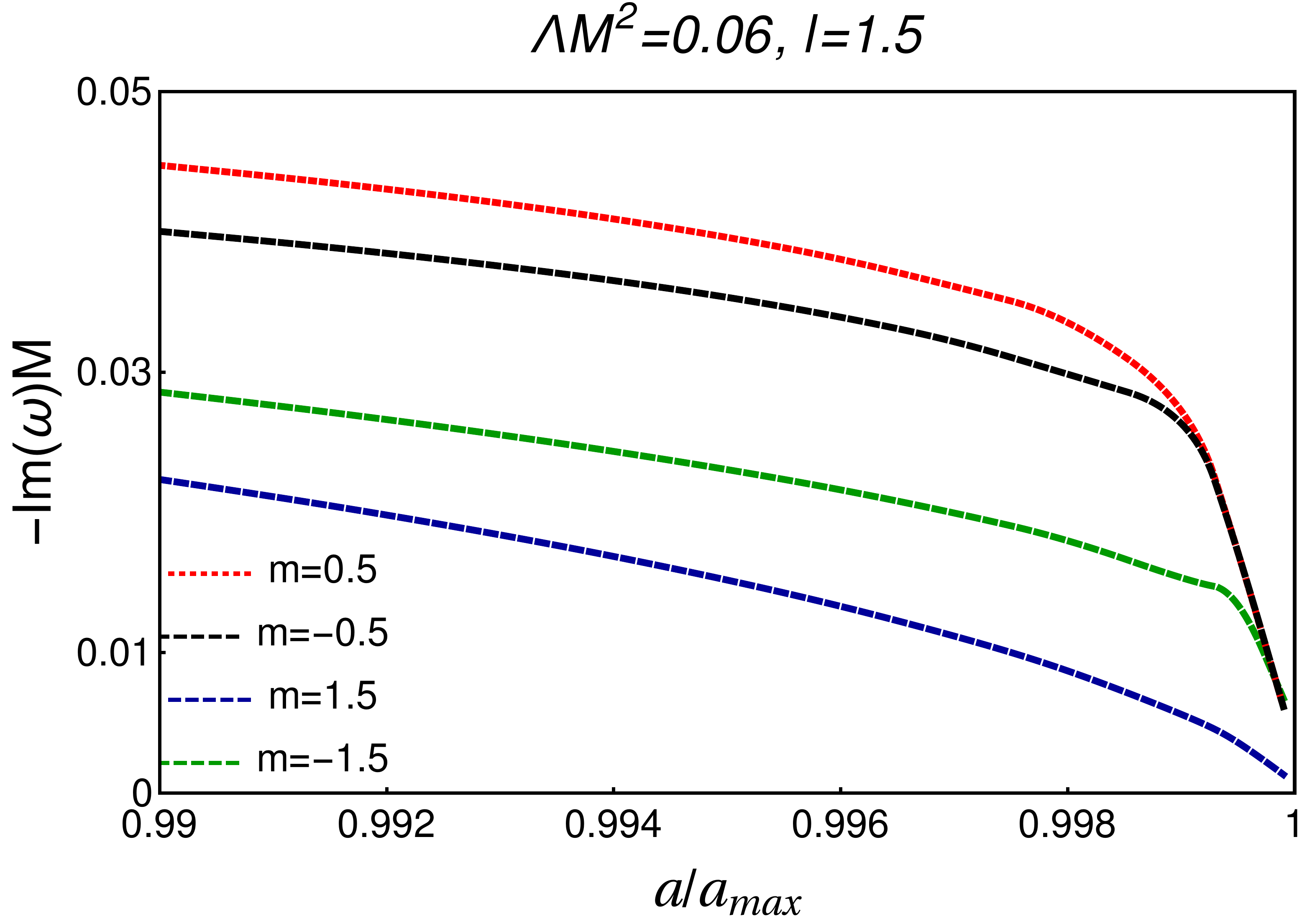}
	\endminipage\hfill
	%%%%%%%%%%%%%%%%%%%%%%%%%
	\minipage{0.33\textwidth}%
	\includegraphics[width=\linewidth]{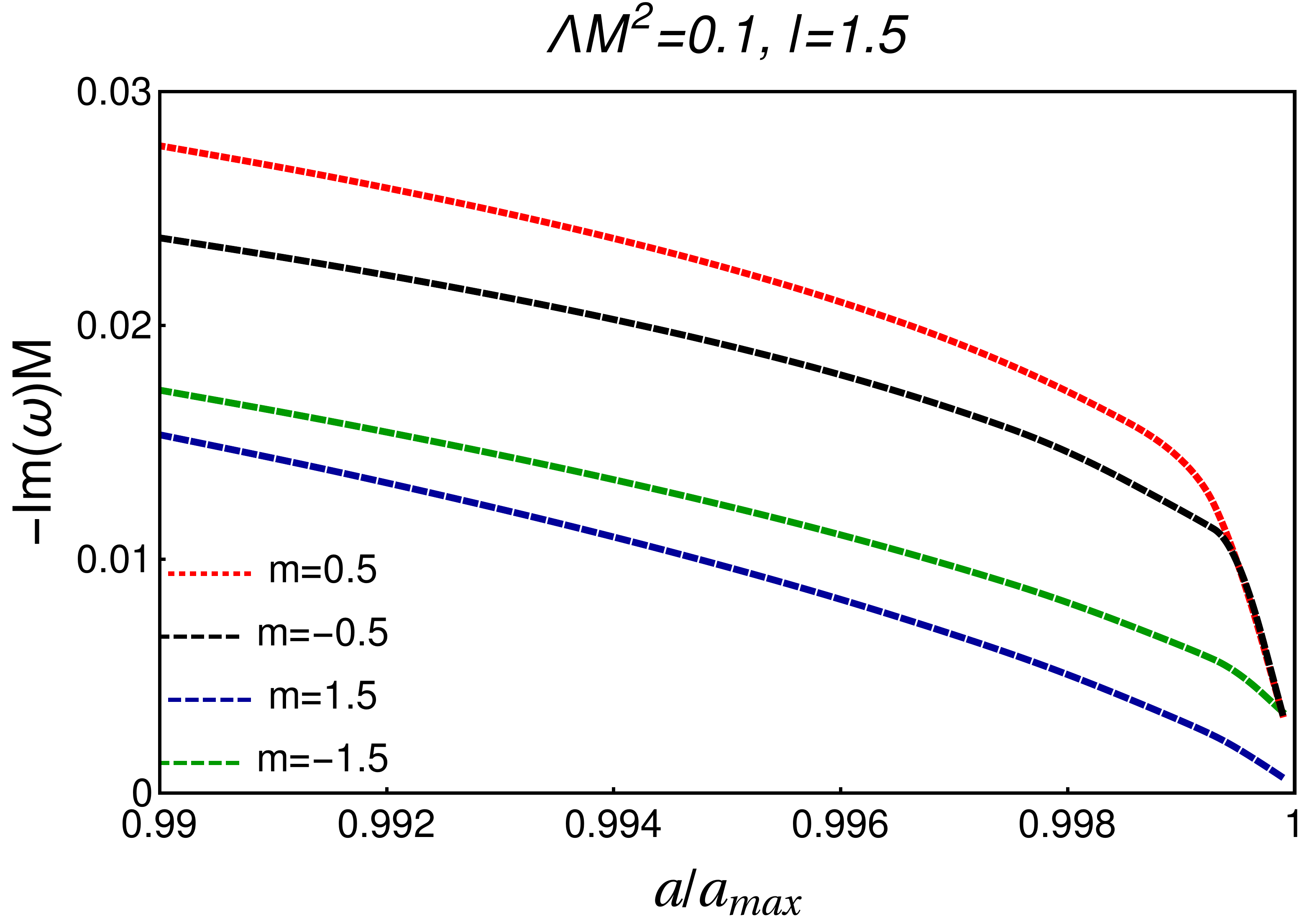}
	\endminipage\hfill
	%%%%%%%%%%%%%%%%%%%%%%%%%
	\minipage{0.33\textwidth}
	\includegraphics[width=\linewidth]{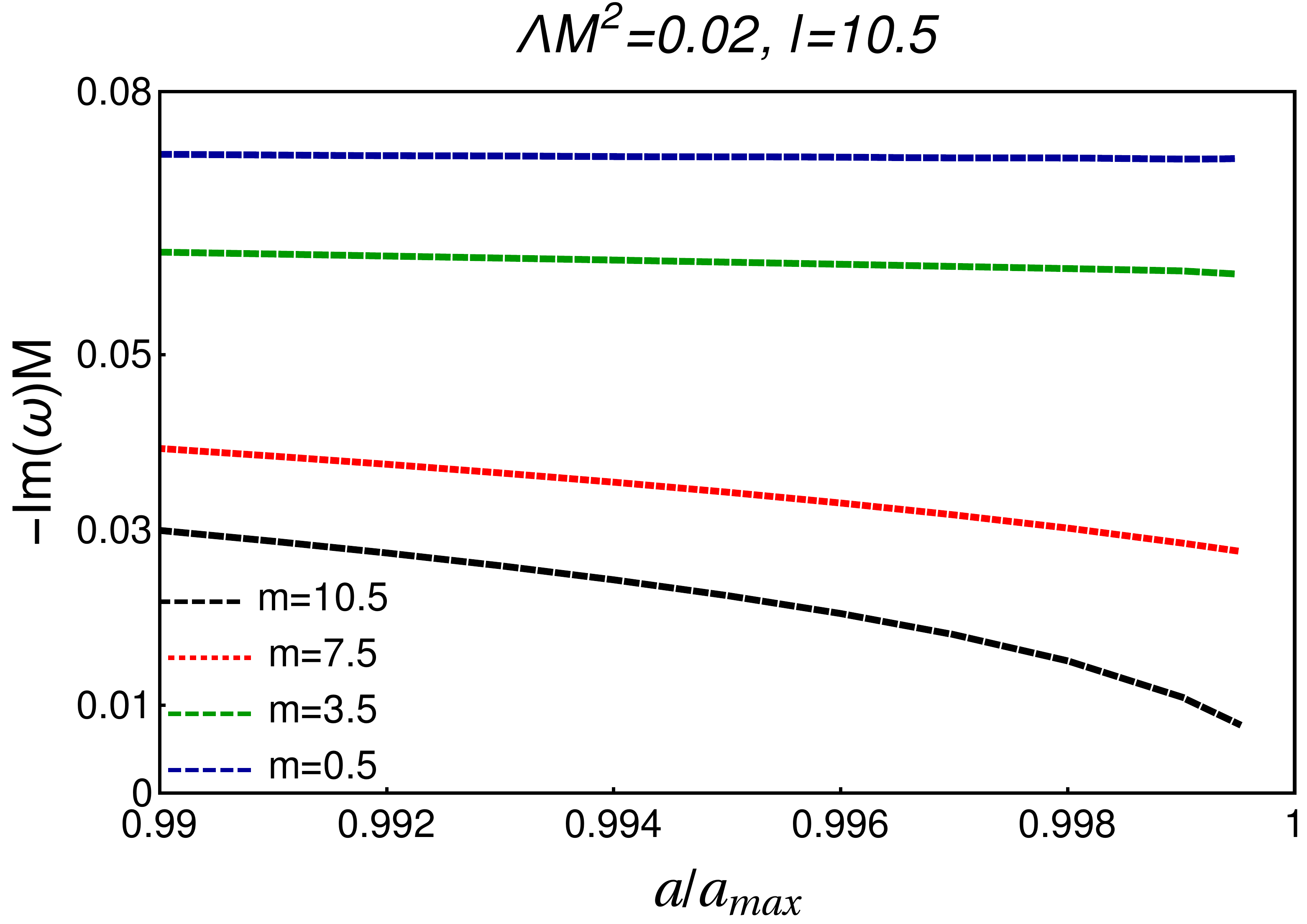}
	\endminipage\hfill
	%%%%%%%%%%%%%%%%%%%%%%%%%
	\minipage{0.33\textwidth}
	\includegraphics[width=\linewidth]{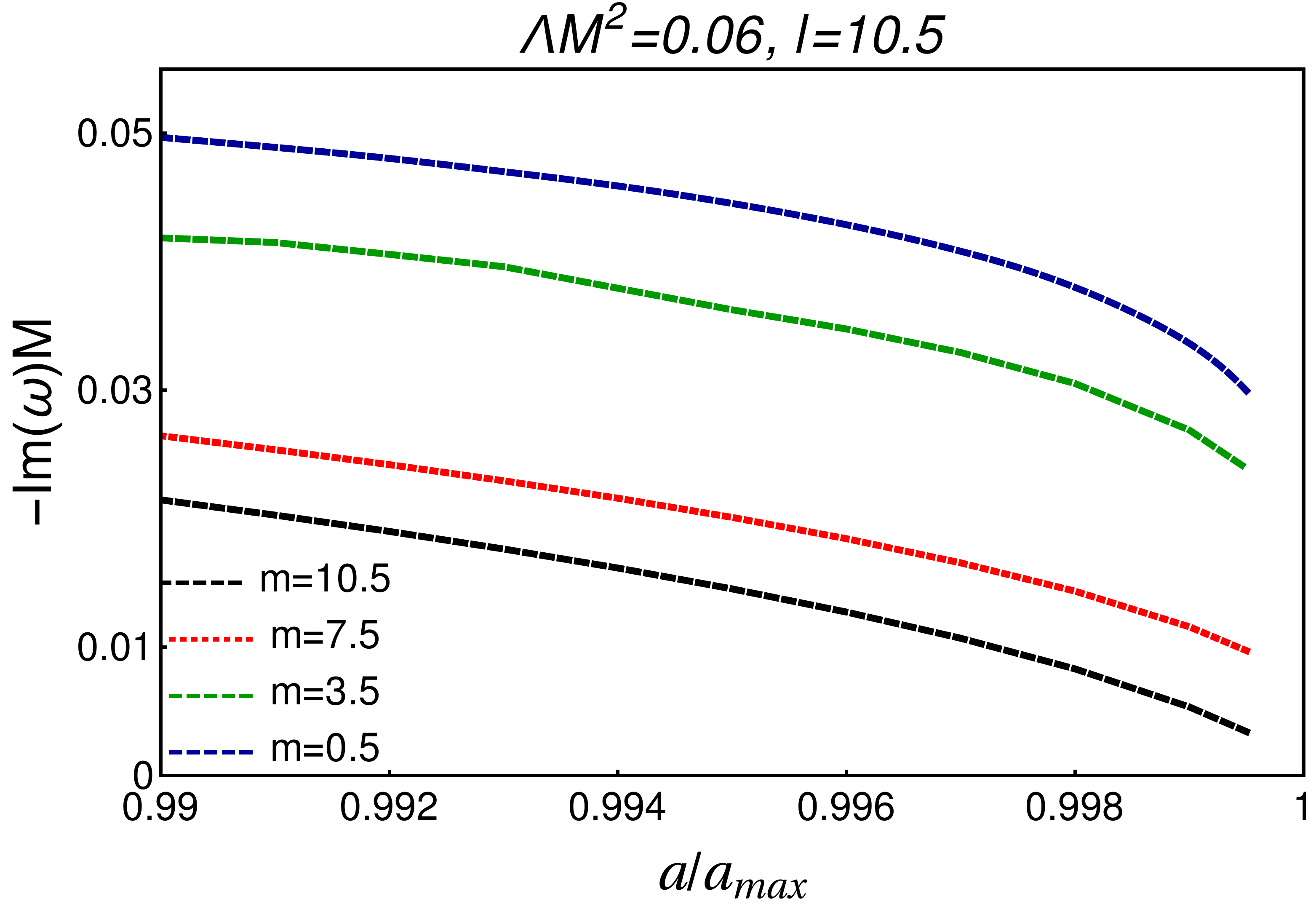}
	\endminipage\hfill
	%%%%%%%%%%%%%%%%%%%%%%%%%
	\minipage{0.33\textwidth}%
	\includegraphics[width=\linewidth]{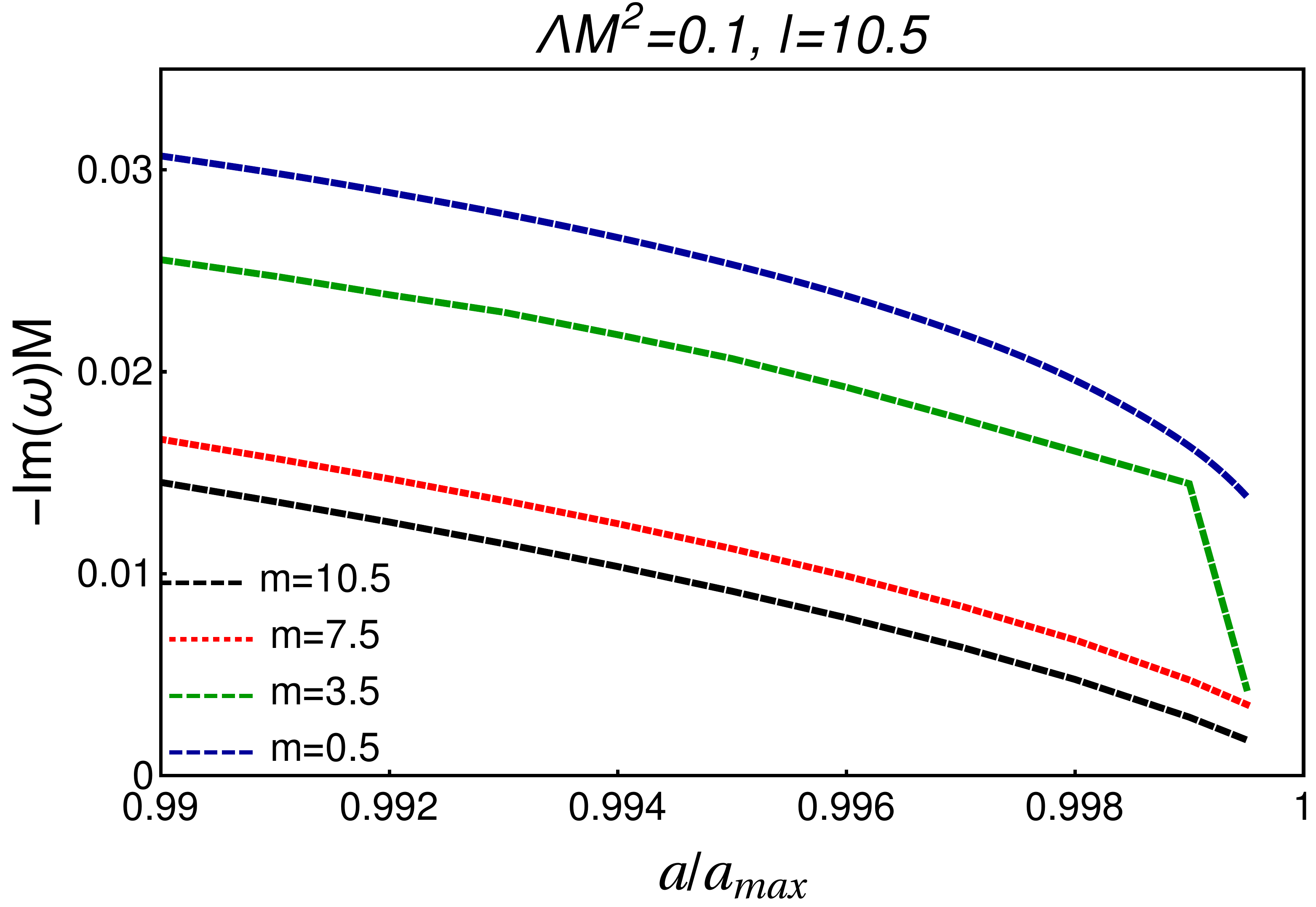}
	\endminipage
	%%%%%%%%%%%%%%%%%%%%%%%%%
	\caption{The variation of imaginary part of quasinormal frequency $-\operatorname{\mathbb{I}m}(\omega)$ as a function of $(a/a_{\rm max})$ has been presented for different values of $\Lambda M^{2}$ and angular eigen mode number $l$. Here, $a_{\rm max}$ denotes the extremal value of the rotation parameter.  As evident from each of these plots, the lowest lying modes always correspond to the mode with $m=l$ regardless of the the values of $\Lambda M^{2}$.}\label{fig_omega_with_m}
\end{figure*}
%%%%%%%%%%%%%%%%%%%%%%%%%%%%%%%%%%%%%%%%%%%%%%%%%%%%%%%%%%%%%%%%%%%%%%%%%%%%%%%%%%%%%%%%%%%%%%%%%%%
%%%%%%%%%%%%%%%%%%%%%%%%%%%%%%%%%%%%%%%%%%%%%%%%%%%%%%%%%%%%%%%%%%%%%%%%%%%%%%%%%%%%%%%%%%%%%%%%%%%
\section{Strong Cosmic Censorship Conjecture for Kerr-de Sitter black holes in presence of Dirac field}\label{main_result}

\begin{figure*}
	\centering 
	%\captionsetup{justification=raggedright}
	\includegraphics[width=0.35\linewidth, height=0.2\textheight]{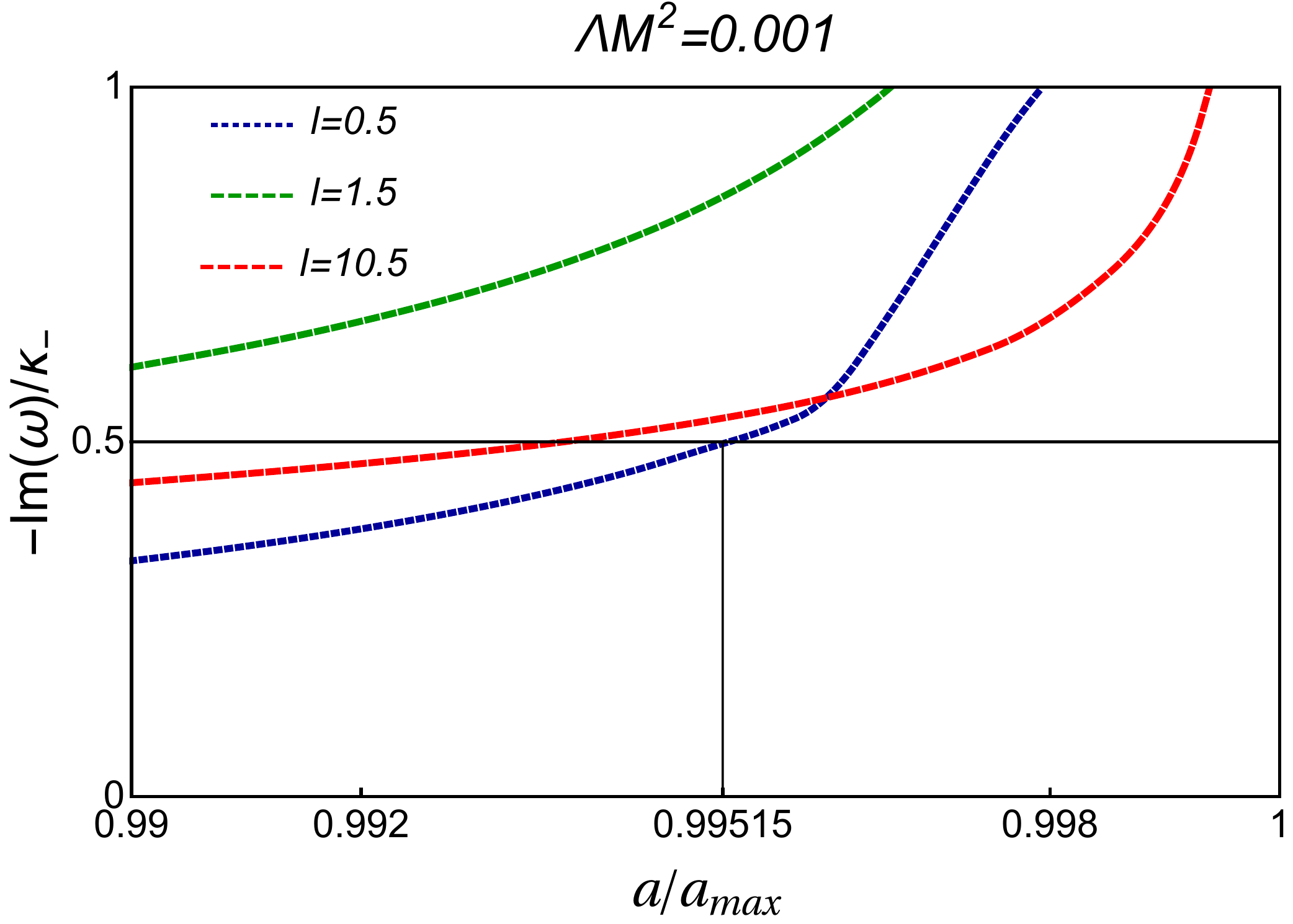} \hspace{1.0cm}\quad
	\includegraphics[width=0.35\linewidth, height=0.2\textheight]{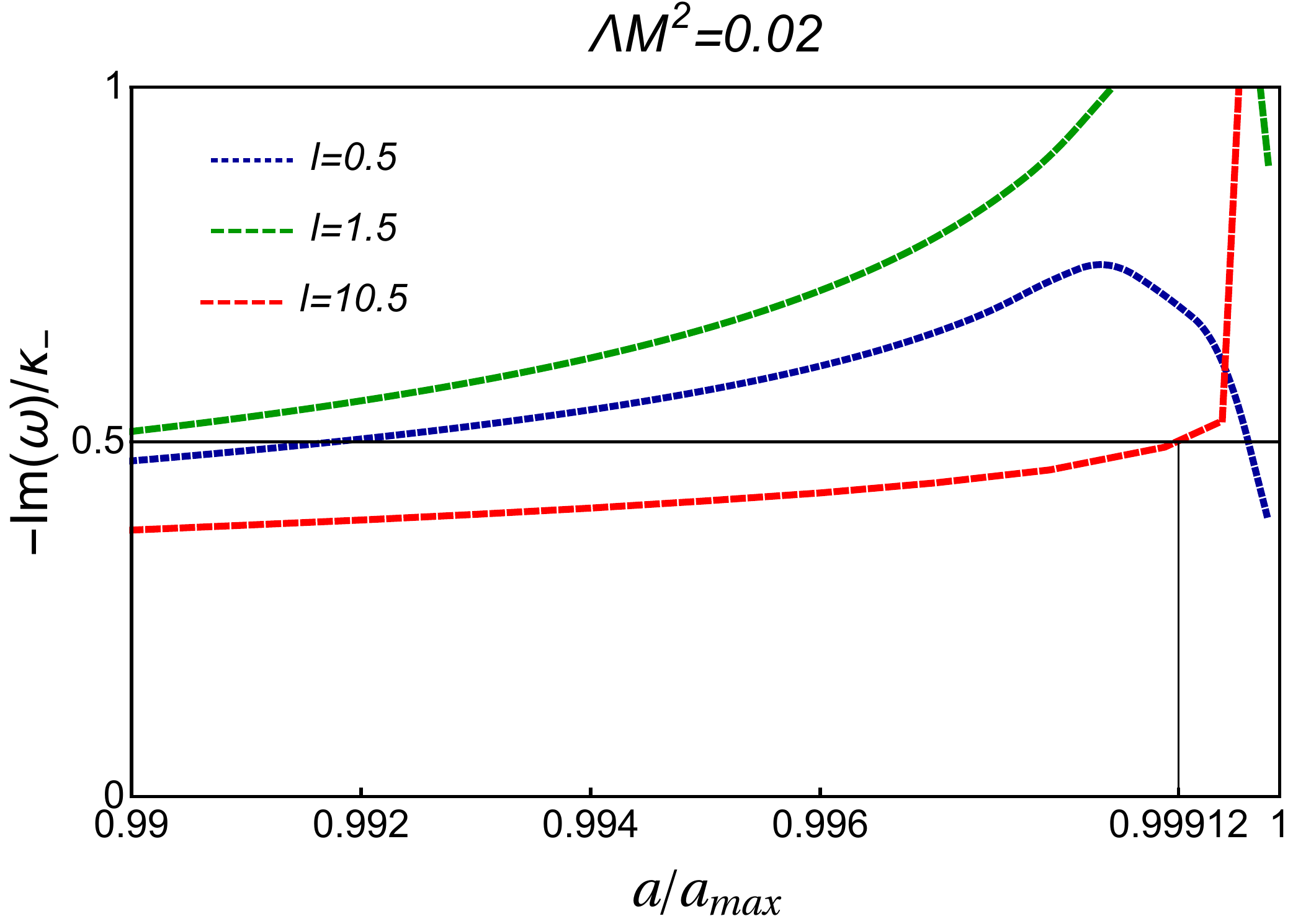}\par\medskip
	\includegraphics[width=0.35\linewidth, height=0.2\textheight]{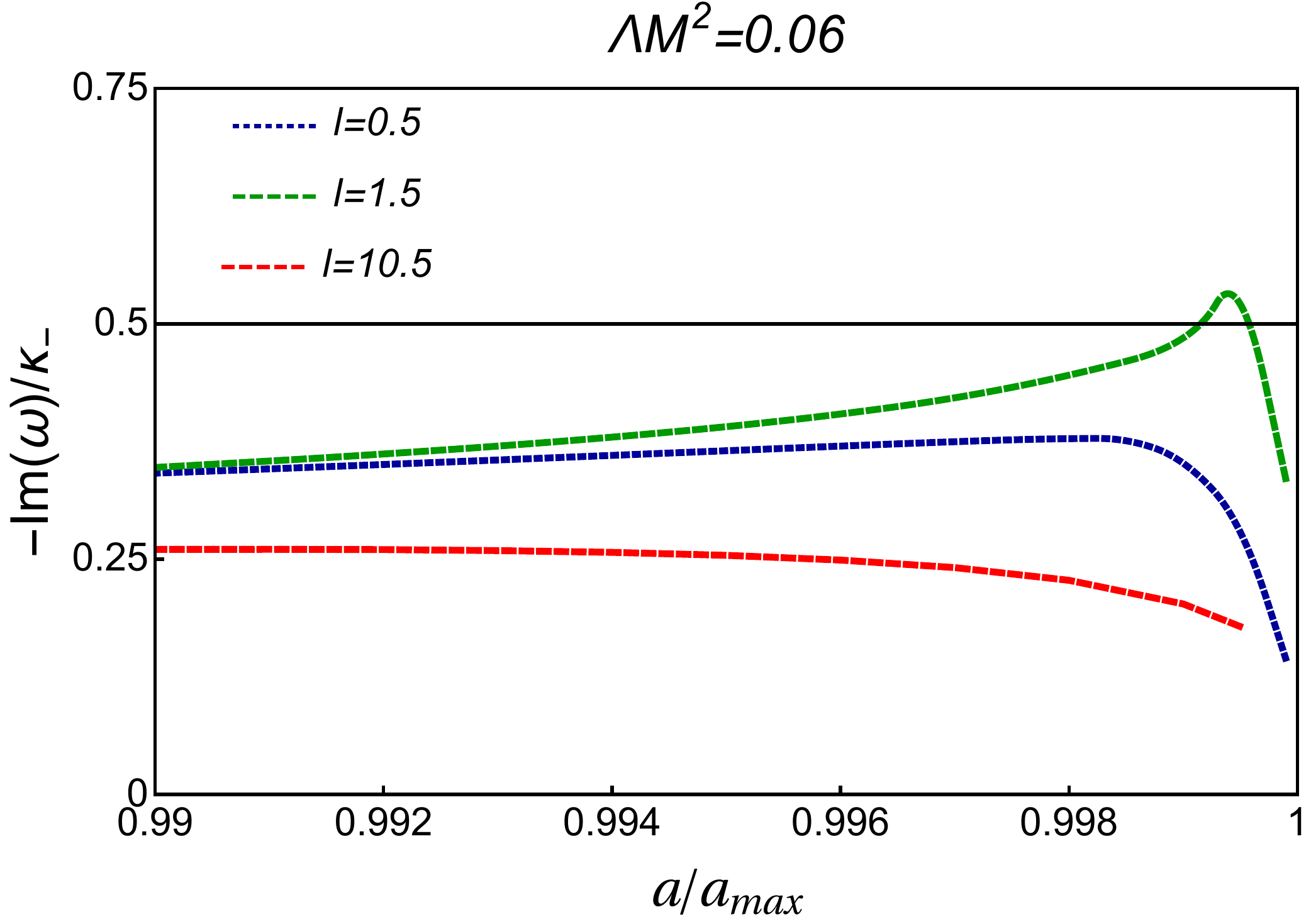} \hspace{1.0cm}
	\includegraphics[width=0.35\linewidth, height=0.2\textheight]{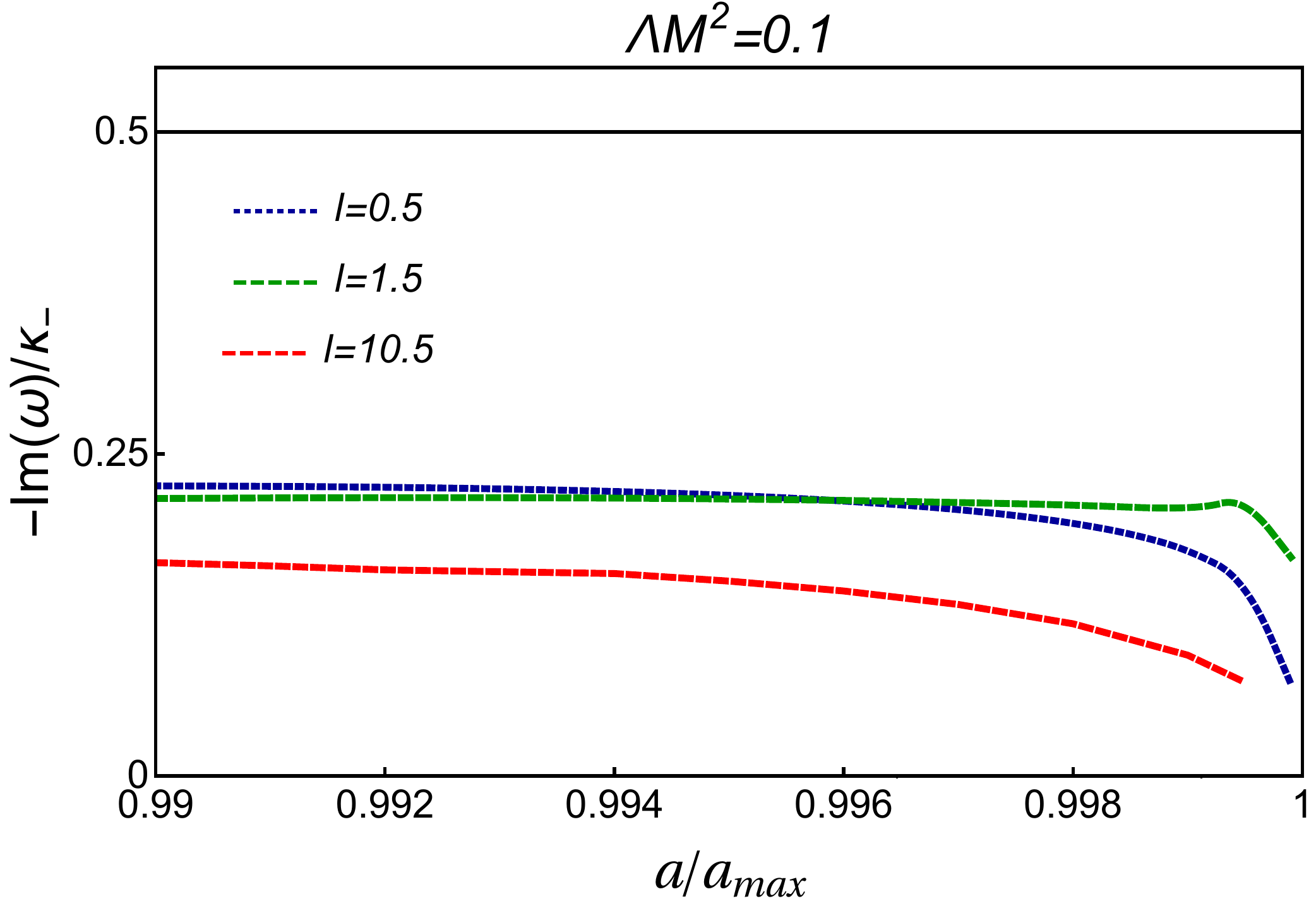}
	\hfill
	%%%%%%%%%%%%%%%%%%%%%%%%
	\caption{ The variation of $Im(\omega)/\kappa_{-}$ as a function of $(a/a_{\rm max})$  has been presented for different values of $\Lambda M^{2}$. The parameter $\beta$ corresponds to the least damped modes of the quasi-normal spectrum. Here, we have considered only the modes with $m=l$. In each of these plots, red, green and blue curves corresponds to value of $Im(\omega)/\kappa_{-}$ for $l=0.5$, $l=1.5$  and $l=10.5$ respectively. The vertical lines in the plots presented in upper-left and upper-right corner corresponds to the value of rotation parameter $a$ for which \SCCC\ gets violated. As evident from the plots, for smaller values of $\Lambda M^{2}$, we can find a parameter space which does not respect the  the conjecture.}\label{fig_SCC_Dirac}
\end{figure*}

In this section, we compute the quasi-normal modes of Kerr-\dS\ black hole in presence of Dirac field using numerical methods. But before that let us put the radial and angular equations given by Eq.~(\ref{Rad_eqn}) and Eq.~(\ref{Ang_eqn}) in standard Teukolsky form which read as follows\cite{Suzuki:1998vy}
%%%%%%%%%%%%%%%%%%%%%%%%%%%%%%%%%%%%%%%%%%%%%%%%%%%%
\begin{align}\label{Rad_Teukolsky}
&\Delta _r^{-s}\partial_{r} \bigg[\Delta _r^{s+1}\partial_{r}R_{+}(r)\bigg]+\bigg[4 i (\alpha +1) r s \omega +2 (1-\alpha ) s\\\nonumber
&-\frac{2 \alpha  r^2 (s+1) (2 s+1)}{a^2}+\frac{V(V-i s \partial_{r}\Delta_{r})}{\Delta _r}-\lambda ^2\bigg]R_{+}(r)=0
\end{align}
%%%%%%%%%%%%%%%%%%%%%%%%%%%%%%%%%%%%%%%%%%%%%%%%%%%%
and
%%%%%%%%%%%%%%%%%%%%%%%%%%%%%%%%%%%%%%%%%%%%%%%%%%%%
\begin{align}\label{Ang_Teukolsky}
&\bigg[\partial_{x}(1+\alpha x^{2})(1-x^{2})\partial_{x}+\lambda^{2}-s(1-\alpha)+\frac{(1+\alpha)^{2}}{\alpha}\xi^{2}\nonumber\\
&-2\alpha x^{2}+\frac{1+\alpha}{1+\alpha x^{2}}\Big\{2s\Big(\alpha m-(1+\alpha)\xi\Big)x-\frac{(1+\alpha)^{2}}{\alpha}\xi^{2}\nonumber\\
&-2m(1+\alpha)\xi+s^{2}\Big\}-\frac{(1+\alpha)^{2}m^{2}}{(1+\alpha x^{2})(1-x^{2})}\nonumber\\
&-\frac{(1+\alpha)(s^{2}+2s m x)}{1-x^{2}}\bigg]S_{+}(x)=0
\end{align}
%%%%%%%%%%%%%%%%%%%%%%%%%%%%%%%%%%%%%%%%%%%%%%%%%%%%

where, $x=\cos\theta$, $s=1/2$ and $\xi=a\omega$. Note that, in the non-rotating limit $(a\to0)$, the angular Teukolsky equation Eq.~(\ref{Ang_Teukolsky}) gives the separation constant as $\lambda\to l(l-1)-s^{2}+s$. This can be used to define the angular eigen mode number $l$ which satisfies the following relation, $l\geq \mathrm{max}(|m|,|s|)$ \cite{PhysRevD.81.044005}. These transformed equations allow us to use the method developed by \cite{Suzuki:1998vy}, who showed that Eq.~(\ref{Ang_Teukolsky}) can be transformed into Heun's equation which, in turn, give us a three-term recurrence relation for the angular equation. This three term recurrence relation can be rewritten in terms of a continued fraction equation which we denote by $P_{1}(\lambda,\omega)=0$.\\
The radial Teukolsky equation Eq.~(\ref{Rad_Teukolsky}) has five regular singularities at $r_+$, $r_-$, $r_{c}$, $-(r_{+}+r_{-}+r_{c})$ and spatial infinity. The quasi-normal modes are defined as the eigen  values of $\omega$ with $\Psi$ satisfying the following boundary condition : there are only outgoing waves at the cosmological horizon $r_{c}$ and only ingoing waves at the event horizon $r_{+}$. In order to satisfy the boundary condition, we write $R_{+}(r)$ as the multiplication of a function $y(z)$ which is regular at the boundary and a factor which is divergent at $r_{+}$ and $r_{c}$, i.e.,
%%%%%%%%%%%%%%%%%%%%%%%%%%%%%%%%%%%%%%%%%%%%%%%%%%%%%%%%%%%%%%%%%%%%%%%%%%%%%%%%%%%%%%%%%%%%%%%%%%%Hence, $\widetilde{R}_{+}(r)$ can be written as the multiplication a Frobenius  series with a factor which is is divergent at $r_{+}$ and $r_{c}$, which for our case of interest becomes \cite{Konoplya:2007zx}
%%%%%%%%%%%%%%%%%%%%%%%%%%%%%%%%%%%%%%%%%%%%%%%%%%%%
\begin{equation}\label{Frobenious_series}
R_{+}(r)=r^{-(2s+1)}\bigg(\frac{r-r_{-}}{r-r_{+}}\bigg)^{s+2i\frac{V(r_{+})}{\Delta_{r}'(r_{+})}}e^{iB(r)}~y(z)
\end{equation}
%%%%%%%%%%%%%%%%%%%%%%%%%%%%%%%%%%%%%%%%%%%%%%%%%%%%
where, $dB/dr=V(r)/\Delta_{r}$ and $z=(r-r_{+})(r_{c}-r_{-})/(r-r_{-})(r_{c}-r_{+})$.  By inserting Eq.~(\ref{Frobenious_series}) into Eq.~(\ref{Rad_Teukolsky}) and expressing $y(z)$ as a Frobenious series of the form $\sum_{n=0}^{\infty}a_{n}z^{n}$, we get a seven-term recurrence relation. Due to convergence problem of seven term recurrence relation \cite{PhysRevD.41.2986}, it is better to reduce the seven-term recurrence relation into a three term recurrence relation using \textit{Gaussian elimination}  method \cite{Konoplya:2011qq, Zhidenko:2009zx}. Similar to the angular equation, this three term recurrence relation can also be written as an infinite continued fraction equation which we denote by $P_{2}(\lambda,\omega)=0$. By solving the angular and radial continued fraction equations simultaneously, one gets the desired quasi-normal modes. However, instead of using the Gaussian elimination procedure to find the three- term recurrence relation for the radial equation, we have employed the \textit{Mathematica} package developed by Jansen \cite{Jansen:2017oag} to find the quasinormal modes. Here, we have solved the angular equation $P_{1}(\lambda,\omega)=0$ and radial equation for $y(z)$ iteratively, taking the approximate value of $\lambda$ in Ref.~\cite{Suzuki:1998vy} as the initial guess value.
%%%%%%%%%%%%%%%%%%%%%%%%%%%%%%%%%%%%%%%%%%%%%%%%%%%%%%%%%%%%%%%%%%%%%%%%%%%%%%%%%%%%%%%%%%%%%%%%%%%(we have also used the \textit{Mathematica} package developed by Jansen to recheck the results \cite{Jansen:2017oag}). 
\\
%%%%%%%%%%%%%%%%%%%%%%%%%%%%%%%%%%%%%%%%%%%%%%%%%%%%%%%%%%%%%%%%%%%%%%%%%%%%%%%%%%%%%%%%%%%%%%%%%%%
%%%%%%%%%%%%%%%%%%%%%%%%%%%%%%%%%%%%%%%%%%%%%%%%%%%%%%%%%%%%%%%%%%%%%%%%%%%%%%%%%%%%%%%%%%%%%%%%%%%
In order to understand the effect of Dirac particles on \SCCC, we need to look for modes for which the parameter $\beta\equiv-\operatorname{\mathbb{I}m}(\omega)/\kappa_{-}$ becomes greater than 1/2. As stated earlier, we are interested in the dominant modes (least damped modes) of the quasinormal spectrum for a given value of angular eigen mode number $l$ for the calculation of the parameter $\beta$. From Fig.~\ref{fig_omega_with_m}, we can see that the dominant mode is always corresponds to modes with $m=l$ for any values of mass scaled cosmological constant $\Lambda M^{2}$ \cite{Konoplya:2007zx, PhysRevD.97.104060}. Moreover, the imaginary part of the quasinormal frequency decreases with the increase of both the rotational parameter $a$ and the cosmological constant $\Lambda$. In this regard, our result is fully consistent with the results presented in \cite{Konoplya:2007zx}. Moreover, near to the extremity, a rapid decrease of the value of $\operatorname{\mathbb{I}m}(\omega)$ is observed. \\
%%%%%%%%%%%%%%%%%%%%%%%%%%%%%%%%%%%%%%%%%%%%%%%%%%%%%%%%%%%%%%%%%%%%%%%%%%%%%%%%%%%%%%%%%%%%%%%%%%%
The variation of the quantity $\operatorname{\mathbb{I}m}(\omega)/\kappa_{-}$ with respect to the rotational parameter $a$ for different values of $\Lambda M^{2}$ is presented in Fig.~\ref{fig_SCC_Dirac}. Here, we consider only the least damped modes for a given value of angular eigen mode number i.e. the modes with  $m=l$. It is interesting to see that for smaller values of mass scaled cosmological constant ($\Lambda M^{2}\approx \mathcal{O}(10^{-3})$), modes corresponding to $l=1/2$ dominate the quasinormal spectrum for certain values of $a$. However, for larger values of $\Lambda M^{2}$,
the eikonal modes (corresponding to the modes with large $l$ value)  become dominant. As evident from Fig.~\ref{fig_SCC_Dirac}, \SCCC\ is respected for larger values of $\Lambda M^{2}$. But for smaller values of mass scaled cosmological constant ($\Lambda M^{2}\lessapprox 0.02$), we are able to find a parameter space for which \SCC\ gets violated. Moreover, in this parameter space, the violation is more severe for smaller values of $\Lambda M^{2}$ since $\beta$ becomes greater than $1/2$ for smaller values of rotational parameter $a$. So the assurance that \SCC\ is always respected in astrophysical black holes \cite{PhysRevD.97.104060, Rahman:2018oso} is no longer valid in the presence of Dirac fields.
%%%%%%%%%%%%%%%%%%%%%%%%%%%%%%%%%%%%%%%%%%%%%%%% \\
%%%%%%%%%%%%%%%%%%%%%%%%%%%%%%%%%%%%%%%%%%%%%%%%%%%%%%%%%%%%%%%%%%%%%%%%%%%%%%%%%%%%%%%%%%%%%%%%%%%
%%%%%%%%%%%%%%%%%%%%%%%%%%%%%%%%%%%%%%%%%%%%%%%%%%%%%%%%%%%%%%%%%%%%%%%%%%%%%%%%%%%%%%%%%%%%%%%%%%%
%%%%%%%%%%%%%%%%%%%%%%%%%
%%%%%%%%%%%%%%%%%%%%%%%%%%%%%%%%%%%%%%%%%%%%%%%%%%%%%%%%%%%%%%%%%%%%%%%%%%%%%%%%%%%%%%%%%%%%%%%%%%%
%%%%%%%%%%%%%%%%%%%%%%%%%%%%%%%%%%%%%%%%%%%%%%%%%%%%%%%%%%%%%%%%%%%%%%%%%%%%%%%%%%%%%%%%%%%%%%%%%%%
%%%%%%%%%%%%%%%%%%%%%%%%%%%%%%%%%%%%%%%%%%%%%%%%%%%%%%%%%%%%%%%%%%%%%%%%%%%%%%%%%%%%%%%%%%%%%%%%%%%
%%%%%%%%%%%%%%%%%%%%%%%%%%%%%%%%%%%%%%%%%%%%%%%%%%%%%%%%%%%%%%%%%%%%%%%%%%%%%%%%%%%%%%%%%%%%%%%%%%%
%%%%%%%%%%%%%%%%%%%%%%%%%%%%%%%%%%%%%%%%%%%%%%%%%%%%%%%%%%%%%%%%%%%%%%%%%%%%%%%%%%%%%%%%%%%%%%%%%%%

\section{Conclusion}\label{conclusion}

In recent years, several examples are found which suggest a breakdown of determinism in \RN-\dS\ black holes under the influence of several fundamental fields \cite{Cardoso:2017soq,Ge:2018vjq,Destounis:2018qnb,Cardoso:2018nvb, Mo:2018nnu, Rahman:2018oso,PhysRevD.99.064014,Liu:2019lon}. Anyway, astrophysically meaningful Kerr-\dS\ black hole solutions seems to respect the conjecture \cite{PhysRevD.97.104060, Rahman:2018oso}. However, the previous analyses were done considering only the effect of scalar fields. In this paper, we extended the study by considering Dirac fields in Kerr-\dS\ background.\\
%%%%%%%%%%%%%%%%%%%%%%%%%%%%%%%%%%%%%%%%%%%%%%%%%%%%%%%%%%%%%%%%%%%%%%%%%%%%%%%%%%%%%%%%%%%%%%%%%%%
By considering the effect of  linear perturbations on the spacetime metric of our interest, we found that there exist weak solutions of Einstein equation at the \CH, if the parameter $\beta$ defined by Eq.~(\ref{SCC_violation}) become greater than $1/2$ which leads to the violation of \SCCC. Comparing our result with Ref.~\cite{PhysRevD.97.104060}, where the effect of scalar fields are considered, we see that the criteria for \SCC\ violation remains the same in the presence of Dirac fields also. Once this criteria is obtained, we performed detailed numerical computation to find lowest lying quasinormal modes to determine the value of $\beta$ as a function of rotational parameter $a$ for different values of mass scaled cosmological constant $\Lambda M^{2}$. In Fig.~\ref{fig_omega_with_m} and Fig.~\ref{fig_SCC_Dirac}, we presented our main results. From these figures, it is clear that the lowest lying quasinormal modes always correspond to the $m=l$ mode. Moreover, an increased value of rotational parameter results in modes with smaller decay rate. Near to the extremity, a rapid decrease of decay rate is observed. The decay rate also decreases with the increase of $\Lambda M^{2}$. From Fig.~\ref{fig_SCC_Dirac}, it is clear that the value of the parameter $\beta$ always remains smaller than $1/2$ for larger values of $\Lambda M^{2}$. Hence, we can conclude that the \SCCC\ is always respected in ``large'' rotating black holes (black holes with $\Lambda M^{2}\gtrapprox 0.02$). However, for smaller values of $\Lambda M^{2}$, we found a parameter space where $\beta$ becomes larger than $1/2$. This, in turn, implies that in ``smaller'' black holes, Dirac fields can be smoothly extended beyond \CH. Hence, in presence of Dirac fields, even the rotational parameter can not save the \SCCC\ for a certain parameter range. Moreover, for a fixed value of cosmological constant, black holes with smaller masses are more prone to violate the conjecture.\\
%%%%%%%%%%%%%%%%%%%%%%%%%%%%%%%%%%%%%%%%%%%%%%%%%%%%%%%%%%%%%%%%%%%%%%%%%%%%%%%%%%%%%%%%%%%%%%%%%%%
Note that, in our work, we have considered the effect of linear perturbation only. It may be possible that the violation of \SCC\ can be prevented by considering the non-linear or quantum gravitational effects. However, Cardoso et al. have shown that for Einstein-Maxwell-scalar field system, the non-linear perturbations can not save the conjecture \cite{PhysRevD.99.064014}. Moreover, a recent study showed that the violation of this conjecture in $2+1$ dimensional BTZ black holes even when quantum correction terms are added \cite{Dias:2019ery}. However, the studies in $3+1$ dimensional black holes remain inconclusive. In particular, it has shown that the quantum correction can either amplify or suppress the blue shift instability in \RN\ black holes \cite{PhysRevLett.70.13}.  It would be interesting to see the those effects on Einstein-Dirac field system, which we leave for future work. Recently, Dafermos and Shlapentokh-Rothman \cite{Dafermos:2018tha} suggested an interesting proposal that the strong cosmic censorship can still be saved if one starts with rough initial data. This idea is further supported by Ref. \cite{Dias:2018etb}, where the authors have studied a coupled gravitational and electro-magnetic perturbation and showed that in order to save the \SCC , one must consider physically reasonable but slightly less smooth initial data. It will be interesting to see whether this proposal can save the conjecture in presence of Dirac field also. However, this is beyond the scope of this paper.
%%%%%%%%%%%%%%%%%%%%%%%%%%%%%%%%%%%%%%%%%%%%%%%%%%%%%%%%%%%%%%%%%%%%%%%%%%%%%%%%%%%%%%%%%%%%%%%%%%%
%%%%%%%%%%%%%%%%%%%%%%%%%%%%%%%%%%%%%%%%%%%%%%%%%%%%%%%%%%%%%%%%%%%%%%%%%%%%%%%%%%%%%%%%%%%%%%%%%%%
%%%%%%%%%%%%%%%%%%%%%%%%%%%%%%%%%%%%%%%%%%%%%%%%%%%%%%%%%%%%
\section*{Acknowledgements}

The author is grateful to Anjan Ananda Sen for valuable discussions and comments. Research of the author is funded by the INSPIRE Fellowship (Reg.No.DST/INSPIRE/03/2015/003030) from Department of Science and Technology, Government of India.

\bibliographystyle{utphys1}
\bibliography{reference1}

\providecommand{\href}[2]{#2}\begingroup\raggedright\begin{thebibliography}{10}

\bibitem{Abbott:2016blz}
{\bfseries LIGO Scientific, Virgo} Collaboration, B.~P. Abbott {\em et~al.},
  ``{Observation of Gravitational Waves from a Binary Black Hole Merger},''
  \href{http://dx.doi.org/10.1103/PhysRevLett.116.061102}{{\em Phys. Rev.
  Lett.} {\bfseries 116} no.~6, (2016) 061102},
\href{http://arxiv.org/abs/1602.03837}{{\ttfamily arXiv:1602.03837 [gr-qc]}}.
%%CITATION = ARXIV:1602.03837;%%.

\bibitem{Akiyama:2019cqa}
{\bfseries Event Horizon Telescope} Collaboration, K.~Akiyama {\em et~al.},
  ``{First M87 Event Horizon Telescope Results. I. The Shadow of the
  Supermassive Black Hole},''
\href{http://dx.doi.org/10.3847/2041-8213/ab0ec7}{{\em Astrophys. J.}
  {\bfseries 875} no.~1, (2019) L1}.
%%CITATION = ASJOA,875,L1;%%.

\bibitem{KONOPLYA2016350}
R.~Konoplya and A.~Zhidenko, ``Detection of gravitational waves from black
  holes: Is there a window for alternative theories?,''
  \href{http://dx.doi.org/https://doi.org/10.1016/j.physletb.2016.03.044}{{\em
  Physics Letters B} {\bfseries 756} (2016) 350 -- 353}.
  \url{http://www.sciencedirect.com/science/article/pii/S0370269316300156}.

\bibitem{Konoplya:2016hmd}
R.~A. Konoplya and A.~Zhidenko, ``{Wormholes versus black holes: quasinormal
  ringing at early and late times},''
  \href{http://dx.doi.org/10.1088/1475-7516/2016/12/043}{{\em JCAP} {\bfseries
  1612} no.~12, (2016) 043},
\href{http://arxiv.org/abs/1606.00517}{{\ttfamily arXiv:1606.00517 [gr-qc]}}.
%%CITATION = ARXIV:1606.00517;%%.

\bibitem{Yunes:2016jcc}
N.~Yunes, K.~Yagi, and F.~Pretorius, ``{Theoretical Physics Implications of the
  Binary Black-Hole Mergers GW150914 and GW151226},''
  \href{http://dx.doi.org/10.1103/PhysRevD.94.084002}{{\em Phys. Rev.}
  {\bfseries D94} no.~8, (2016) 084002},
\href{http://arxiv.org/abs/1603.08955}{{\ttfamily arXiv:1603.08955 [gr-qc]}}.
%%CITATION = ARXIV:1603.08955;%%.

\bibitem{1973IJTP....7..183S}
M.~{Simpson} and R.~{Penrose}, ``{Internal Instability in a
  Reissner-Nordstr{\"o}m Black Hole},''
  \href{http://dx.doi.org/10.1007/BF00792069}{{\em International Journal of
  Theoretical Physics} {\bfseries 7} (Apr., 1973) 183--197}.

\bibitem{PhysRevD.41.1796}
E.~Poisson and W.~Israel, ``Internal structure of black holes,''
  \href{http://dx.doi.org/10.1103/PhysRevD.41.1796}{{\em Phys. Rev. D}
  {\bfseries 41} (Mar, 1990) 1796--1809}.
  \url{https://link.aps.org/doi/10.1103/PhysRevD.41.1796}.

\bibitem{Dafermos:2003wr}
M.~Dafermos, ``{The Interior of charged black holes and the problem of
  uniqueness in general relativity},'' {\em Commun. Pure Appl. Math.}
  {\bfseries 58} (2005) 0445--0504,
\href{http://arxiv.org/abs/gr-qc/0307013}{{\ttfamily arXiv:gr-qc/0307013
  [gr-qc]}}.
%%CITATION = GR-QC/0307013;%%.

\bibitem{Dafermos:2012np}
M.~Dafermos, ``{Black holes without spacelike singularities},''
  \href{http://dx.doi.org/10.1007/s00220-014-2063-4}{{\em Commun. Math. Phys.}
  {\bfseries 332} (2014) 729--757},
\href{http://arxiv.org/abs/1201.1797}{{\ttfamily arXiv:1201.1797 [gr-qc]}}.
%%CITATION = ARXIV:1201.1797;%%.

\bibitem{Costa:2017tjc}
J.~L. Costa, P.~M. Girão, J.~Natário, and J.~D. Silva, ``{On the Occurrence
  of Mass Inflation for the Einstein–Maxwell-Scalar Field System with a
  Cosmological Constant and an Exponential Price Law},''
  \href{http://dx.doi.org/10.1007/s00220-018-3122-z}{{\em Commun. Math. Phys.}
  {\bfseries 361} no.~1, (2018) 289--341},
\href{http://arxiv.org/abs/1707.08975}{{\ttfamily arXiv:1707.08975 [gr-qc]}}.
%%CITATION = ARXIV:1707.08975;%%.

\bibitem{Costa:2014yha}
J.~L. Costa, P.~M. Girão, J.~Natário, and J.~D. Silva, ``{On the global
  uniqueness for the Einstein-Maxwell-scalar field system with a cosmological
  constant: I. Well posedness and breakdown criterion},''
  \href{http://dx.doi.org/10.1088/0264-9381/32/1/015017}{{\em Class. Quant.
  Grav.} {\bfseries 32} no.~1, (2015) 015017},
\href{http://arxiv.org/abs/1406.7245}{{\ttfamily arXiv:1406.7245 [gr-qc]}}.
%%CITATION = ARXIV:1406.7245;%%.

\bibitem{0264-9381-16-12A-302}
D.~Christodoulou, ``On the global initial value problem and the issue of
  singularities,'' {\em Classical and Quantum Gravity} {\bfseries 16} no.~12A,
  (1999) A23. \url{http://stacks.iop.org/0264-9381/16/i=12A/a=302}.

\bibitem{Christodoulou:2008nj}
D.~Christodoulou, \href{http://dx.doi.org/10.1142/9789814374552_0002}{``{The
  Formation of Black Holes in General Relativity},''} in {\em {On recent
  developments in theoretical and experimental general relativity, astrophysics
  and relativistic field theories. Proceedings, 12th Marcel Grossmann Meeting
  on General Relativity, Paris, France, July 12-18, 2009. Vol. 1-3}},
  pp.~24--34.
\newblock 2008.
\newblock
\href{http://arxiv.org/abs/0805.3880}{{\ttfamily arXiv:0805.3880 [gr-qc]}}.
\newblock
%%CITATION = ARXIV:0805.3880;%%.

\bibitem{PhysRevD.97.104060}
O.~J.~C. Dias, F.~C. Eperon, H.~S. Reall, and J.~E. Santos, ``Strong cosmic
  censorship in de sitter space,''
  \href{http://dx.doi.org/10.1103/PhysRevD.97.104060}{{\em Phys. Rev. D}
  {\bfseries 97} (May, 2018) 104060}.
  \url{https://link.aps.org/doi/10.1103/PhysRevD.97.104060}.

\bibitem{Dafermos:2018tha}
M.~Dafermos and Y.~Shlapentokh-Rothman, ``{Rough initial data and the strength
  of the blue-shift instability on cosmological black holes with $\Lambda >
  0$},'' \href{http://dx.doi.org/10.1088/1361-6382/aadbcf}{{\em Class. Quant.
  Grav.} {\bfseries 35} no.~19, (2018) 195010},
\href{http://arxiv.org/abs/1805.08764}{{\ttfamily arXiv:1805.08764 [gr-qc]}}.
%%CITATION = ARXIV:1805.08764;%%.

\bibitem{Luk:2015qja}
J.~Luk and S.-J. Oh, ``{Proof of linear instability of the
  Reissner–Nordström Cauchy horizon under scalar perturbations},''
  \href{http://dx.doi.org/10.1215/00127094-3715189}{{\em Duke Math. J.}
  {\bfseries 166} no.~3, (2017) 437--493},
\href{http://arxiv.org/abs/1501.04598}{{\ttfamily arXiv:1501.04598 [gr-qc]}}.
%%CITATION = ARXIV:1501.04598;%%.

\bibitem{Dafermos:2015bzz}
M.~Dafermos and Y.~Shlapentokh-Rothman, ``{Time-Translation Invariance of
  Scattering Maps and Blue-Shift Instabilities on Kerr Black Hole
  Spacetimes},'' \href{http://dx.doi.org/10.1007/s00220-016-2771-z}{{\em
  Commun. Math. Phys.} {\bfseries 350} no.~3, (2017) 985--1016},
\href{http://arxiv.org/abs/1512.08260}{{\ttfamily arXiv:1512.08260 [gr-qc]}}.
%%CITATION = ARXIV:1512.08260;%%.

\bibitem{Ge:2018vjq}
B.~Ge, J.~Jiang, B.~Wang, H.~Zhang, and Z.~Zhong, ``{Strong cosmic censorship
  for the massless Dirac field in the Reissner-Nordstrom-de Sitter
  spacetime},'' \href{http://dx.doi.org/10.1007/JHEP01(2019)123}{{\em JHEP}
  {\bfseries 01} (2019) 123},
\href{http://arxiv.org/abs/1810.12128}{{\ttfamily arXiv:1810.12128 [gr-qc]}}.
%%CITATION = ARXIV:1810.12128;%%.

\bibitem{Destounis:2018qnb}
K.~Destounis, ``{Charged Fermions and Strong Cosmic Censorship},''
\href{http://arxiv.org/abs/1811.10629}{{\ttfamily arXiv:1811.10629 [gr-qc]}}.
%%CITATION = ARXIV:1811.10629;%%.

\bibitem{Chambers:1997ef}
C.~M. Chambers, ``{The Cauchy horizon in black hole de sitter space-times},''
  {\em Annals Israel Phys. Soc.} {\bfseries 13} (1997) 33,
  \href{http://arxiv.org/abs/gr-qc/9709025}{{\ttfamily arXiv:gr-qc/9709025
  [gr-qc]}}.
[,33(1997)].
%%CITATION = GR-QC/9709025;%%.

\bibitem{Cardoso:2017soq}
V.~Cardoso, J.~L. Costa, K.~Destounis, P.~Hintz, and A.~Jansen, ``{Quasinormal
  modes and Strong Cosmic Censorship},''
  \href{http://dx.doi.org/10.1103/PhysRevLett.120.031103}{{\em Phys. Rev.
  Lett.} {\bfseries 120} no.~3, (2018) 031103},
\href{http://arxiv.org/abs/1711.10502}{{\ttfamily arXiv:1711.10502 [gr-qc]}}.
%%CITATION = ARXIV:1711.10502;%%.

\bibitem{Cardoso:2018nvb}
V.~Cardoso, J.~L. Costa, K.~Destounis, P.~Hintz, and A.~Jansen, ``{Strong
  cosmic censorship in charged black-hole spacetimes: still subtle},''
  \href{http://dx.doi.org/10.1103/PhysRevD.98.104007}{{\em Phys. Rev.}
  {\bfseries D98} no.~10, (2018) 104007},
\href{http://arxiv.org/abs/1808.03631}{{\ttfamily arXiv:1808.03631 [gr-qc]}}.
%%CITATION = ARXIV:1808.03631;%%.

\bibitem{Mo:2018nnu}
Y.~Mo, Y.~Tian, B.~Wang, H.~Zhang, and Z.~Zhong, ``{Strong cosmic censorship
  for the massless charged scalar field in the Reissner-Nordstrom–de Sitter
  spacetime},'' \href{http://dx.doi.org/10.1103/PhysRevD.98.124025}{{\em Phys.
  Rev.} {\bfseries D98} no.~12, (2018) 124025},
\href{http://arxiv.org/abs/1808.03635}{{\ttfamily arXiv:1808.03635 [gr-qc]}}.
%%CITATION = ARXIV:1808.03635;%%.

\bibitem{Rahman:2018oso}
M.~Rahman, S.~Chakraborty, S.~SenGupta, and A.~A. Sen, ``{Fate of Strong Cosmic
  Censorship Conjecture in Presence of Higher Spacetime Dimensions},''
  \href{http://dx.doi.org/10.1007/JHEP03(2019)178}{{\em JHEP} {\bfseries 03}
  (2019) 178},
\href{http://arxiv.org/abs/1811.08538}{{\ttfamily arXiv:1811.08538 [gr-qc]}}.
%%CITATION = ARXIV:1811.08538;%%.

\bibitem{PhysRevD.99.064014}
R.~Luna, M.~Zilh\~ao, V.~Cardoso, J.~a.~L. Costa, and J.~Nat\'ario, ``Strong
  cosmic censorship: The nonlinear story,''
  \href{http://dx.doi.org/10.1103/PhysRevD.99.064014}{{\em Phys. Rev. D}
  {\bfseries 99} (Mar, 2019) 064014}.
  \url{https://link.aps.org/doi/10.1103/PhysRevD.99.064014}.

\bibitem{Liu:2019lon}
H.~Liu, Z.~Tang, K.~Destounis, B.~Wang, E.~Papantonopoulos, and H.~Zhang,
  ``{Strong Cosmic Censorship in higher-dimensional Reissner-Nordström-de
  Sitter spacetime},'' \href{http://dx.doi.org/10.1007/JHEP03(2019)187}{{\em
  JHEP} {\bfseries 03} (2019) 187},
\href{http://arxiv.org/abs/1902.01865}{{\ttfamily arXiv:1902.01865 [gr-qc]}}.
%%CITATION = ARXIV:1902.01865;%%.

\bibitem{Rahman:2020guv}
M.~Rahman, S.~Mitra, and S.~Chakraborty, ``{Strong cosmic censorship conjecture
  with NUT charge and conformal coupling},''
  \href{http://arxiv.org/abs/2001.00599}{{\ttfamily arXiv:2001.00599 [gr-qc]}}.

\bibitem{Suzuki:1998vy}
H.~Suzuki, E.~Takasugi, and H.~Umetsu, ``{Perturbations of Kerr-de Sitter black
  hole and Heun's equations},''
  \href{http://dx.doi.org/10.1143/PTP.100.491}{{\em Prog. Theor. Phys.}
  {\bfseries 100} (1998) 491--505},
\href{http://arxiv.org/abs/gr-qc/9805064}{{\ttfamily arXiv:gr-qc/9805064
  [gr-qc]}}.
%%CITATION = GR-QC/9805064;%%.

\bibitem{Chandrasekhar:1985kt}
S.~Chandrasekhar, ``{The mathematical theory of black holes},'' in {\em
  {Oxford, UK: Clarendon (1992) 646 p., OXFORD, UK: CLARENDON (1985) 646 P.}}
\newblock
1985.
\newblock
%%CITATION = INSPIRE-224457;%%.

\bibitem{Frolov:1998wf}
V.~P. Frolov and I.~D. Novikov, eds.,
  \href{http://dx.doi.org/10.1007/978-94-011-5139-9}{{\em {Black hole physics:
  Basic concepts and new developments}}}, vol.~96.
\newblock
1998.
\newblock
%%CITATION = FTPHD,96,;%%.

\bibitem{PhysRevD.28.1291}
U.~Khanal, ``Rotating black hole in asymptotic de sitter space: Perturbation of
  the space-time with spin fields,''
  \href{http://dx.doi.org/10.1103/PhysRevD.28.1291}{{\em Phys. Rev. D}
  {\bfseries 28} (Sep, 1983) 1291--1297}.
  \url{https://link.aps.org/doi/10.1103/PhysRevD.28.1291}.

\bibitem{Chang:2005ki}
J.-F. Chang and Y.-G. Shen, ``{Neutrino quasinormal modes of a Kerr-Newman-de
  Sitter black hole},''
  \href{http://dx.doi.org/10.1016/j.nuclphysb.2005.01.043}{{\em Nucl. Phys.}
  {\bfseries B712} (2005) 347--370},
\href{http://arxiv.org/abs/gr-qc/0502083}{{\ttfamily arXiv:gr-qc/0502083
  [gr-qc]}}.
%%CITATION = GR-QC/0502083;%%.

\bibitem{Zannias:2017ign}
T.~Zannias, ``{On causality violation on a Kerr-de Sitter spacetime},''
  \href{http://dx.doi.org/10.1007/s10714-018-2456-3}{{\em Gen. Rel. Grav.}
  {\bfseries 50} no.~10, (2018) 134},
\href{http://arxiv.org/abs/1711.01313}{{\ttfamily arXiv:1711.01313 [gr-qc]}}.
%%CITATION = ARXIV:1711.01313;%%.

\bibitem{PhysRevD.81.044005}
S.~Yoshida, N.~Uchikata, and T.~Futamase, ``Quasinormal modes of kerr--de
  sitter black holes,''
  \href{http://dx.doi.org/10.1103/PhysRevD.81.044005}{{\em Phys. Rev. D}
  {\bfseries 81} (Feb, 2010) 044005}.
  \url{https://link.aps.org/doi/10.1103/PhysRevD.81.044005}.

\bibitem{PhysRevD.41.2986}
E.~W. Leaver, ``Quasinormal modes of reissner-nordstr\"om black holes,''
  \href{http://dx.doi.org/10.1103/PhysRevD.41.2986}{{\em Phys. Rev. D}
  {\bfseries 41} (May, 1990) 2986--2997}.
  \url{https://link.aps.org/doi/10.1103/PhysRevD.41.2986}.

\bibitem{Konoplya:2011qq}
R.~A. Konoplya and A.~Zhidenko, ``{Quasinormal modes of black holes: From
  astrophysics to string theory},''
  \href{http://dx.doi.org/10.1103/RevModPhys.83.793}{{\em Rev. Mod. Phys.}
  {\bfseries 83} (2011) 793--836},
\href{http://arxiv.org/abs/1102.4014}{{\ttfamily arXiv:1102.4014 [gr-qc]}}.
%%CITATION = ARXIV:1102.4014;%%.

\bibitem{Zhidenko:2009zx}
A.~Zhidenko, {\em {Linear perturbations of black holes: stability, quasi-normal
  modes and tails}}.
\newblock PhD thesis, Sao Paulo U., 2009.
\newblock \href{http://arxiv.org/abs/0903.3555}{{\ttfamily arXiv:0903.3555
  [gr-qc]}}.
\newblock
\url{https://gwic.ligo.org/thesisprize/2009/Zhidenko_Thesis.pdf}.
\newblock
%%CITATION = ARXIV:0903.3555;%%.

\bibitem{Jansen:2017oag}
A.~Jansen, ``{Overdamped modes in Schwarzschild-de Sitter and a Mathematica
  package for the numerical computation of quasinormal modes},''
  \href{http://dx.doi.org/10.1140/epjp/i2017-11825-9}{{\em Eur. Phys. J. Plus}
  {\bfseries 132} no.~12, (2017) 546},
\href{http://arxiv.org/abs/1709.09178}{{\ttfamily arXiv:1709.09178 [gr-qc]}}.
%%CITATION = ARXIV:1709.09178;%%.

\bibitem{Konoplya:2007zx}
R.~A. Konoplya and A.~Zhidenko, ``{Decay of a charged scalar and Dirac fields
  in the Kerr-Newman-de Sitter background},''
  \href{http://dx.doi.org/10.1103/PhysRevD.76.084018,
  10.1103/PhysRevD.90.029901}{{\em Phys. Rev.} {\bfseries D76} no.~8, (2007)
  084018}, \href{http://arxiv.org/abs/0707.1890}{{\ttfamily arXiv:0707.1890
  [hep-th]}}.
[Erratum: Phys. Rev.D90,no.2,029901(2014)].
%%CITATION = ARXIV:0707.1890;%%.

\bibitem{Dias:2019ery}
O.~J.~C. Dias, H.~S. Reall, and J.~E. Santos, ``{The BTZ black hole violates
  strong cosmic censorship},''
\href{http://arxiv.org/abs/1906.08265}{{\ttfamily arXiv:1906.08265 [hep-th]}}.
%%CITATION = ARXIV:1906.08265;%%.

\bibitem{PhysRevLett.70.13}
R.~Balbinot and E.~Poisson, ``Mass inflation: The semiclassical regime,''
  \href{http://dx.doi.org/10.1103/PhysRevLett.70.13}{{\em Phys. Rev. Lett.}
  {\bfseries 70} (Jan, 1993) 13--16}.
  \url{https://link.aps.org/doi/10.1103/PhysRevLett.70.13}.

\bibitem{Dias:2018etb}
O.~J.~C. Dias, H.~S. Reall, and J.~E. Santos, ``{Strong cosmic censorship:
  taking the rough with the smooth},''
  \href{http://dx.doi.org/10.1007/JHEP10(2018)001}{{\em JHEP} {\bfseries 10}
  (2018) 001},
\href{http://arxiv.org/abs/1808.02895}{{\ttfamily arXiv:1808.02895 [gr-qc]}}.
%%CITATION = ARXIV:1808.02895;%%.

\end{thebibliography}\endgroup
%\tableofcontents
\end{document}